\documentclass[]{style/nseJournal}

\usepackage[utf8]{inputenc}
\usepackage[T1]{fontenc}
\usepackage{amssymb}
\usepackage{amsmath}
\usepackage{comment}
\usepackage{booktabs, adjustbox} 
\usepackage{subcaption} 
\usepackage{multirow} 
\usepackage[section]{placeins}
\usepackage{layouts}
\usepackage{xcolor}
\begin{document}

\title{Generational variance reduction in Monte Carlo criticality simulations as a way of mitigating unwanted correlations}

\addAuthor{\correspondingAuthor{Kévin Fröhlicher}}{a}
\correspondingEmail{kevin.frohlicher@irsn.fr}
\addAuthor{Eric Dumonteil}{b}
\addAuthor{Loïc Thulliez}{b}
\addAuthor{Julien Taforeau}{a}
\addAuthor{Mariya Brovchenko}{a}

\addAffiliation{a}{Institut de Radioprotection et de Sûreté Nucléaire (IRSN)\\ 31 avenue de la Division Leclerc, 92260, Fontenay-aux-Roses, France}
\addAffiliation{b}{IRFU, CEA, Université Paris-Saclay\\ 91191, Gif-sur-Yvette, France}

\addKeyword{Monte Carlo Criticality Simulations}
\addKeyword{Neutron Clustering}
\addKeyword{Power Iteration}
\addKeyword{Variance Reduction}
\addKeyword{Adaptive Multilevel Splitting}

\titlePage
\begin{abstract}

Monte Carlo criticality simulations are widely used in nuclear safety demonstrations, as they offer an arbitrarily precise estimation of global and local tallies while making very few assumptions.
However, since the inception of such numerical approaches, it is well known that bias might affect both the estimation of errors on these tallies and the tallies themselves.
In particular, stochastic modeling approaches developed in the past decade have shed light on the prominent role played by spatial correlations through a phenomenon called neutron clustering.
This effect is particularly of great significance when simulating loosely coupled systems (i.e., with a high dominance ratio). 
In order to tackle this problem, this paper proposes to recast the power iteration technique of Monte Carlo criticality codes into a variance reduction technique called Adaptative Multilevel Splitting.
The central idea is that iterating over neutron generations can be seen as pushing a sub-population of neutrons towards a generational detector (instead of a spatial detector as variance reduction techniques usually do).
While both approaches allow for neutron population control, the former blindly removes or splits neutrons.
In contrast, the latter optimizes spatial, generational, and spectral attributes of neutrons when they are removed or split through an adjoint flux estimation, hence tempering both generational and spatial correlations.
This is illustrated in the present article with a simple case of a bare slab reactor in the one speed theory on which the Adaptive Multilevel Splitting was applied and compared to variations of the Monte Carlo power iteration method used in neutron transport.
Besides looking at the resulting efficiency of the methods, this work also aims at highlighting the main mechanisms of the Adaptive Multilevel Splitting in criticality calculations.

\end{abstract}

\section{Introduction}\label{sec:introduction}

For a long time, dating back 80 years, the simulation of neutron transport in multiplicative media has been one of the main motivations leading the development of intensive systems (digital) calculation capabilities and the Monte Carlo algorithm itself. 
Nuclear engineers and researchers still consider Monte Carlo methods as high fidelity methods, compared to deterministic ones, since they estimate global and local tallies with an arbitrary precision while making very few hypotheses.
Nuclear data uncertainties are often considered as the only source of uncertainty (while others such as technological uncertainties are usually neglected or ignored), aside from the stochastic fluctuations intrinsic to the very nature of the method.
Therefore, Monte Carlo simulations are widely used as reference calculations when validating new models/methods.


Criticality calculations have been used for decades in reactor physics to characterize the behavior of multiplicative nuclear systems through their $k_\textrm{eff}$ by solving the transport critical equation \cite{dickinson1976monte,goad1959monte,mendelson1968monte,MOORE1976THEMETHODS,rief1967reactor}. 
This equation takes the form of an eigenvalue equation in which fixed neutron sources are neglected (for instance, spontaneous fissions), and the production term of the equation is modified to ensure that the neutron population remains constant through generations.
Solving this $k$-eigenvalue equation by Monte Carlo methods is generally done using an iterative algorithm based on the power iteration method, and therefore allows characterizing the fundamental mode of the system as if it was exactly critical.
Despite fundamental questions related to the inner nature of the problem that is solved due to the renormalization of fission neutrons by the $k_\textrm{eff}$ \cite{cullen2003static}, this method has made a consensus when it comes to criticality calculations.
It is now widely used not only in nuclear criticality safety but also in reactor physics.
For loosely coupled systems, however, it can exhibit convergence issues \cite{Dumonteil2010DominanceSystems} that can lead to potentially significant errors in estimating global and local tallies and their statistical uncertainties.
By loosely coupled systems, we mean systems in which neutrons may have difficulty travelling from one part of the geometry to another over several generations, typically large systems. 


Indeed, in the late 60's, different works shed light on the biases on the $k_\textrm{eff}$ estimation in criticality calculations \cite{osti_4835730,osti_6720467,macmillan1973monte,gelbard1974monte,brissenden1986biases} which were however increasingly used by the nuclear industry.
Later, Ueki et al. \cite{ueki2003autocorrelation}, and Dumonteil et al. \cite{dumonteil2014particle} highlighted respectively the fact that generational and spatial correlations were also a source of biases in the spatial flux estimation.
Additionally, different works \cite{lux1991monte} pointed out that tallies estimators are usually built from observations drawn in successive generations of neutrons which lead to an actual underestimation bias of the tallies uncertainty \cite{Ueki1996ErrorCalculations,ueki2003autocorrelation}.


At the heart of these observations lies the fact that, while both types of correlations (generational and spatial) are deeply rooted in the fission phenomenon and therefore also develop in actual nuclear configurations \cite{dumonteil2021patchy}, their magnitude might indeed soar in numerical simulations due to the combination of the population control and the small number of neutrons that can be simulated compared to natural systems.
In particular, the so-called neutron clustering effect has received considerable attention in the past decade.
It is typical of branching spatial processes since its origin is found in the asymmetry between neutron captures, which occur everywhere, and neutron births, which can only happen in the vicinity of other neutrons.
Whenever the system is loosely coupled, and even in the presence of absorbing boundaries \cite{Zoia2014ClusteringGeometries}, the asymmetry creates spatial patterns of randomly distributed neutron clusters.
This results in the under-sampling of some regions and ultimately leads to biased estimates of the global (e.g., $k_\textrm{eff}$) and local (e.g., flux) tallies \cite{Miao2018PredictingProcess}, which can have drastic consequences when feedback effects are taken into account through multi-physics coupling \cite{cosgrove2020neutron,Cosgrove2021CounteringInjection}. 
This phenomenon, called neutron clustering, has been investigated using statistical mechanics tools and, in particular, could be modeled using the so-called branching Brownian motion, which couples a Galton-Watson birth-death process to standard Brownian motion \cite{dumonteil2014particle, Zoia2014ClusteringGeometries, mulatier2015random, dumonteil2017clustering}.
Although neutron clustering is usually mitigated by sampling more particles into the Monte Carlo simulation, different strategies have been tested to avoid the occurrence of this phenomenon.
While attenuating the phenomenon, neither the introduction of two-time scales that reproduce the effect of delayed neutrons nor the presence or absence of population control \cite{DeMulatier2015TheRevisited} affect this qualitative picture \cite{Houchmandzadeh2015NeutronDelayed,Bonnet2022SpaceEvents}.
Because the persistence of neutron families was the key to counter this effect in simulations, beneficial modifications of the Monte Carlo power iteration method were recently investigated \cite{Sutton2022TowardClustering,cosgrove2020neutron,Mickus2021DoesCalculations}.


Starting from these last observations, this paper proposes a different approach to tackle generational and spatial correlations (hence, to temper Monte Carlo criticality biases).
The observation that drives our approach is that the power iteration randomly kills or splits neutrons during population control. The only way to ensure that neutron families extinction is slowed down is to restart neutrons that will survive for many generations.
Hence, the paradigm change here consists of seeing the population control acting on a super/subcritical medium as a way to either select neutrons or enforce neutrons persistence through generations.
In other words, the goal is to estimate the asymptotic state of our system conditioned on its survival: such approaches are known in mathematics as Fleming-Viot processes \cite{fleming1979some,asselah2016fleming}. 
These processes can also be seen as an estimator for rare events \cite{Cerou2019OnSplitting} and therefore as a variance reduction techniques \cite{dunn2011exploring} which primarily aims at "pushing" neutron to a given "detector" without introducing any bias in the estimates of tallies associated to this detector.
These techniques can use an importance map whose quality will condition the improvement in the method efficiency. In the present case, our detector could be a generational detector, and the importance map could be the adjoint flux \cite{ussachoff1956equation,hurwitz1964physical,lewins1965importance} that could be estimated on-the-fly or using external codes.
A generalization of such Fleming-Viot processes that allows for handling importance functions has recently appeared.
This method is based on the use of particle splitting and using an on-the-fly estimation of the importance levels at which particles are split \cite{cerou2007adaptive}.
It has been named Adaptative Multilevel Splitting (AMS) and has been adapted to neutron transport in the context of shielding calculations \cite{Louvin2017AdaptiveTransport,Brun2015TRIPOLI-4Code}.
The present paper will show that modification of this variance reduction technique that has proven successful in shielding calculations can also help mitigate correlations and biases in Monte Carlo criticality calculations.
It will also show that AMS can be used alone or on top of other population control techniques (such as the branchless collision method \cite{lux1991monte}).


The paper is organized as follows.
In Section \ref{sec:AMS}, the Adaptive Multilevel Splitting method, and its extension to criticality calculations are presented, while Section \ref{sec:results} outlines the main numerical results and discussions about methods performances.
All methods were compared regarding $k_\textrm{eff}$ and flux estimations, as well as their impact on spatial and generational correlations.

\FloatBarrier
\section{Adaptive Multilevel Splitting}\label{sec:AMS}

\subsection{Original algorithm for particle transport}

The Adaptive Multilevel Splitting (AMS) is a method initially developed in applied mathematics to compute rare events probability.
Initially intended for continuous Markov chains \cite{cerou2007adaptive}, it was then adapted to discrete Markov chains \cite{brehier2016unbiasedness} and used in particle transport for attenuation and radiation protection problems \cite{louvin2017adaptive, louvin2017development}.

\begin{figure*}[!htb]
 \centering
  \includegraphics[width=1\textwidth]{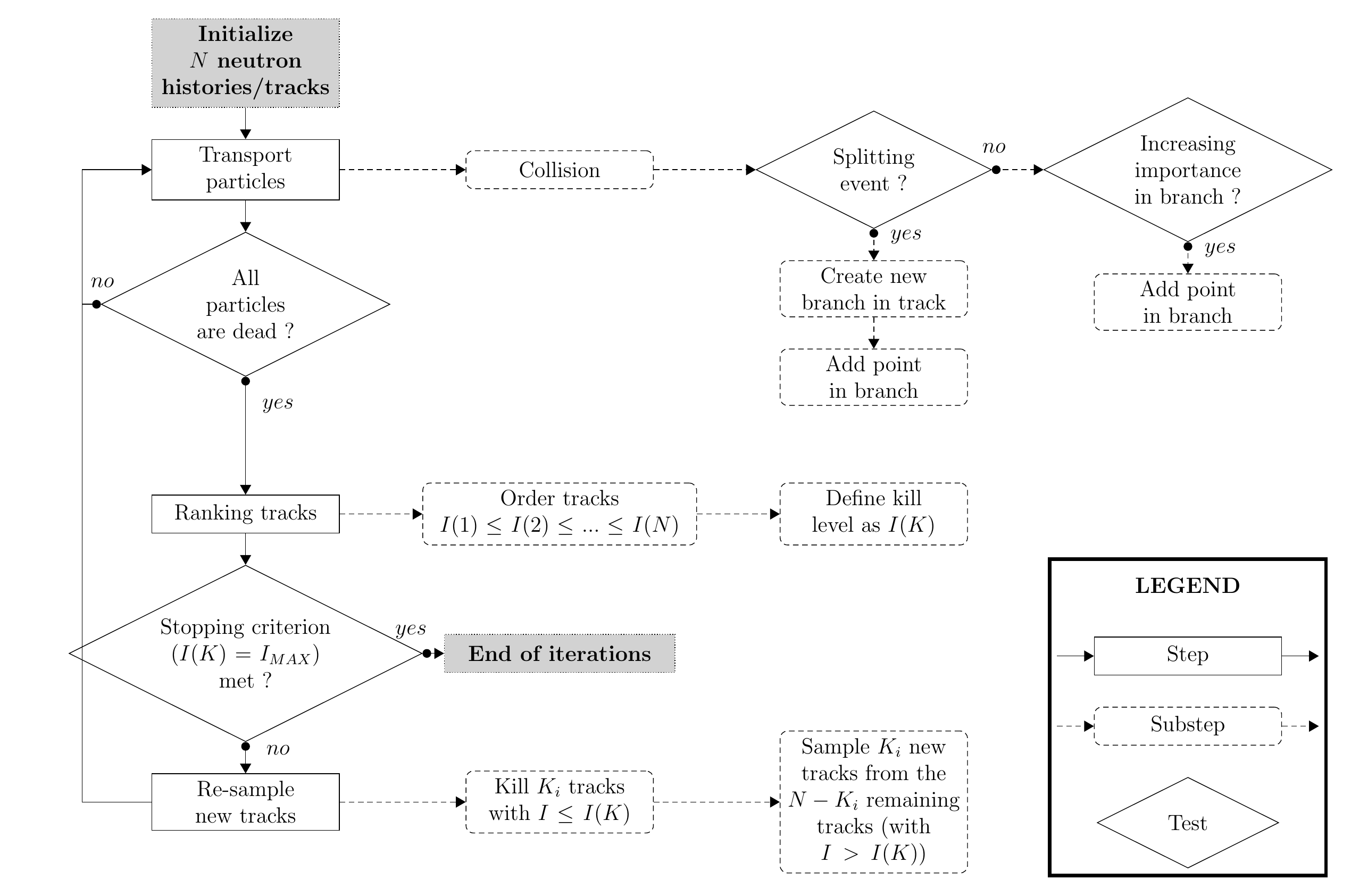}
 \caption{AMS algorithm}
 \label{fig:AMS_flowchart}
\end{figure*}

The concept is to re-sample particles towards a detector iteratively.
The key idea underlying the method is to re-sample particle histories closer and closer to a detector.
To this aim, neutron histories are first simulated from birth to death (disappearance of all its particles by capture, leakage, Russian Roulette) and ranked following an importance criterion.
According to the ranking, the least important histories are deleted, and histories re-sampled among the remaining ones.
The general algorithm is illustrated in Figure \ref{fig:AMS_flowchart}.
The general algorithm is here described for analog transport of particles, it is however possible to use the AMS in a weighted Monte Carlo game \cite{brehier2016unbiasedness}.

\subsubsection{AMS tree structure and transport step}\label{sec:AMS_transport_step}
\begin{figure*}[!htb]
 \centering
  \includegraphics[width=1\textwidth]{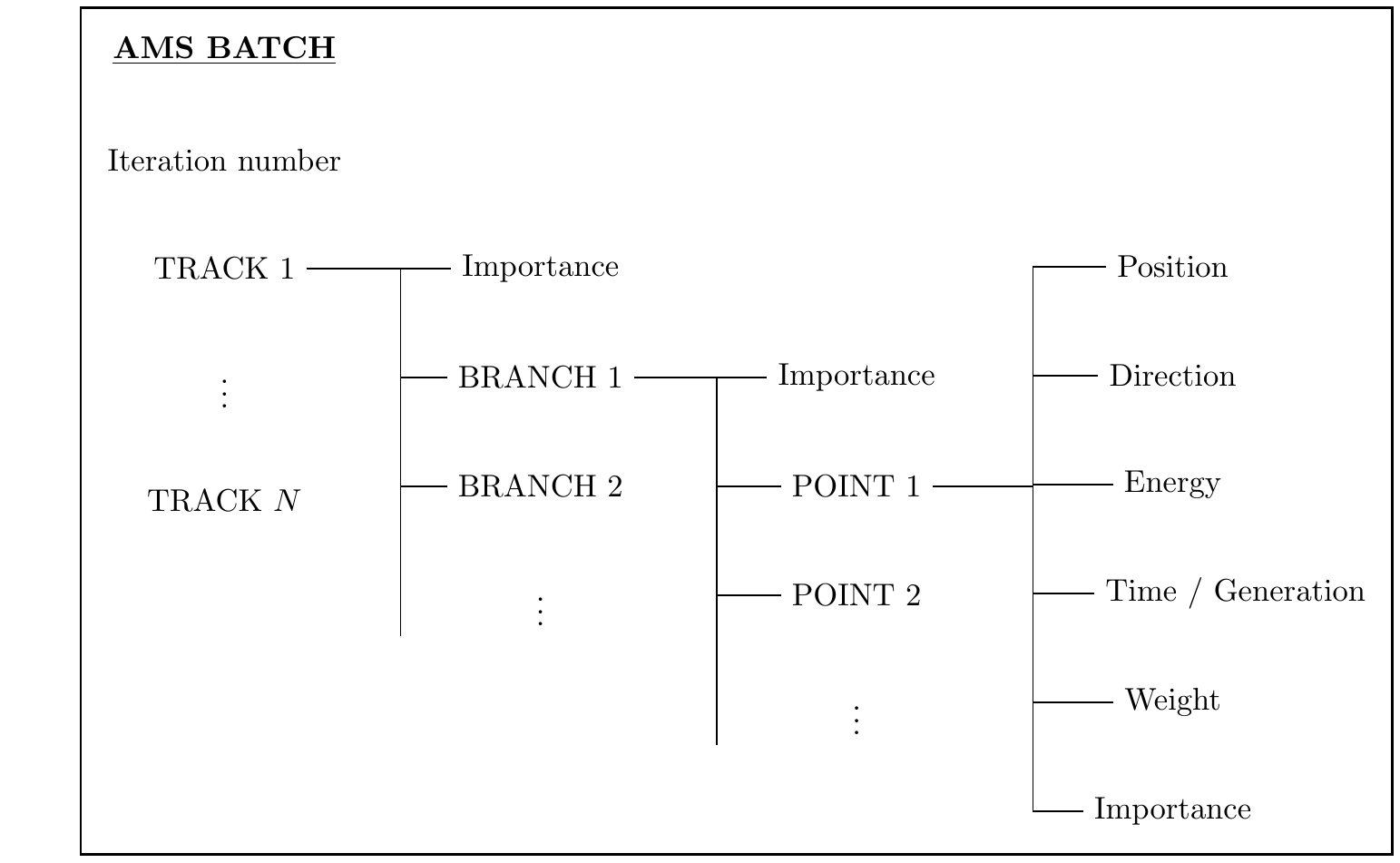}
 \caption{AMS track/branch/point structure}
 \label{fig:AMS_structure}
\end{figure*}

The AMS consists of successive fixed source simulations, where each batch , i.e. each simulation, is composed of $N$ tracks (initially independent) representing $N$ particle histories.
Here, a particle history is the whole trajectory of a particle and its progeny arising from collisions and splitting events, from the birth of the initial particle to the death of all its progeny.
At each collision, the outgoing particle of the collision is assigned an importance value using a cost function (see Section \ref{sec:AMS_importance_function}).
If the particle importance is higher than the previous branch point importance, the particle is saved as a point in the AMS structure.
Each track initially starts with a unique branch, and new branches are appended every time a splitting event occurs (it can be physical like fission or numerical like splitting in a weighted Monte Carlo game).
At any time, a track importance is equal to the maximum importance of its branches, and the importance of a branch is equal to the maximum importance amongst its points.
The resulting tree structure is illustrated in Figure \ref{fig:AMS_structure}.

\subsubsection{Ranking the histories}\label{sec:AMS_ranking_step}
Once all particles in an iteration $i$ are dead (by capture, leakage, Russian Roulette), the $N$ tracks are ranked in increasing order of importance.
At this point, a kill level $I_\textrm{kill}$ is defined by the importance of the $K$-th worst track, where $K$ is a user-defined parameter.
\begin{equation}\label{eq:kill_level}
    I_\textrm{kill}^{(i)} \equiv I^{(i)}(K)
\end{equation}
When a track has reached the detector, its importance is set to infinity.
At the end of an iteration, the algorithm samples new tracks according to the following description if the stopping criterion that is defined hereinafter has not been met.

AMS iterations stop when 
\begin{equation}\label{eq:stopping_criterion}
    I_\textrm{kill}^{(i)} = I^{(i)}_{MAX}
\end{equation}
where $I^{(i)}_{MAX}$ is the maximum track importance at iteration $i$.
If the algorithm iterated correctly (see discussion on the importance in Section \ref{sec:AMS_importance_function}), it should stop when
\begin{equation}
    I_\textrm{kill}^{(i)} = I^{(i)}(K) = \infty.
\end{equation}
This equation implies that tracks $K$ to $N$ have reached the detector (since their importance is superior to $I^i(K)$), hence, at least $N-K+1$ tracks have reached the detector.

\subsubsection{Sampling new particles}\label{sec:AMS_sampling_step}

After the kill level has been computed, all tracks whose importance is lower or equal to this level are deleted from the batch structure.
Since multiple tracks can have the same importance, the number of deleted tracks is not necessarily equal to $K$ (it can be higher), and we denote it $K_i$ at iteration $i$, where $K_i \equiv card(S) \geq K$, with $S$ defined as
\begin{equation}\label{eq:tracks_deleted}
    S = \left\{ k,\ k\in \left[1;N\right]\ |\ \ I^{(i)}(k)\leq I_\textrm{kill} \right\}.
\end{equation}

\begin{figure*}[!htb]
 \centering
 \begin{subfigure}[t]{0.48\textwidth}
  \centering
  \includegraphics[width=1.1\textwidth]{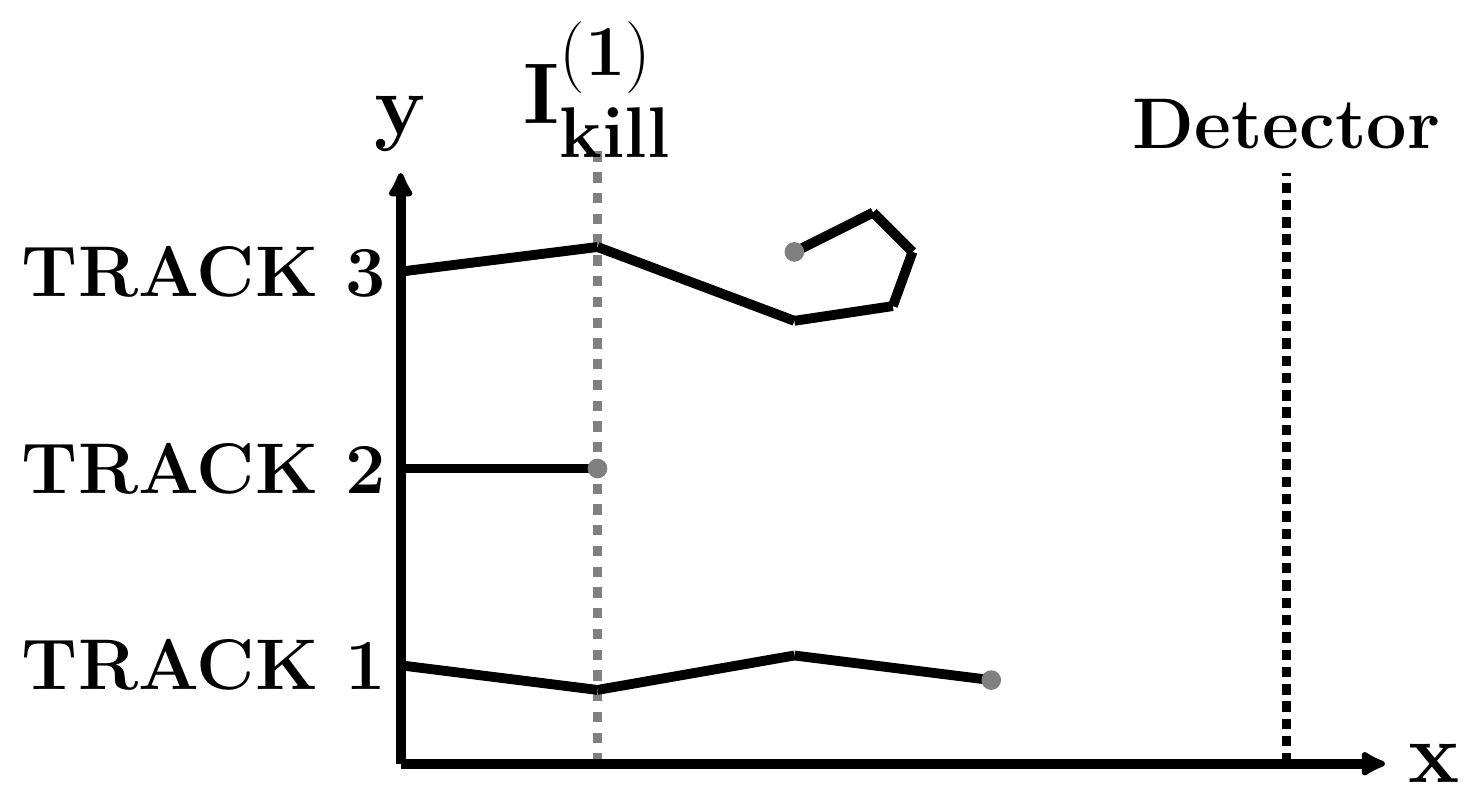}
  \subcaption{Iteration 1}
  \label{fig:AMS_iteration_1}
 \end{subfigure}
 \hfill
 \begin{subfigure}[t]{0.48\textwidth}
  \centering
  \includegraphics[width=1.1\textwidth]{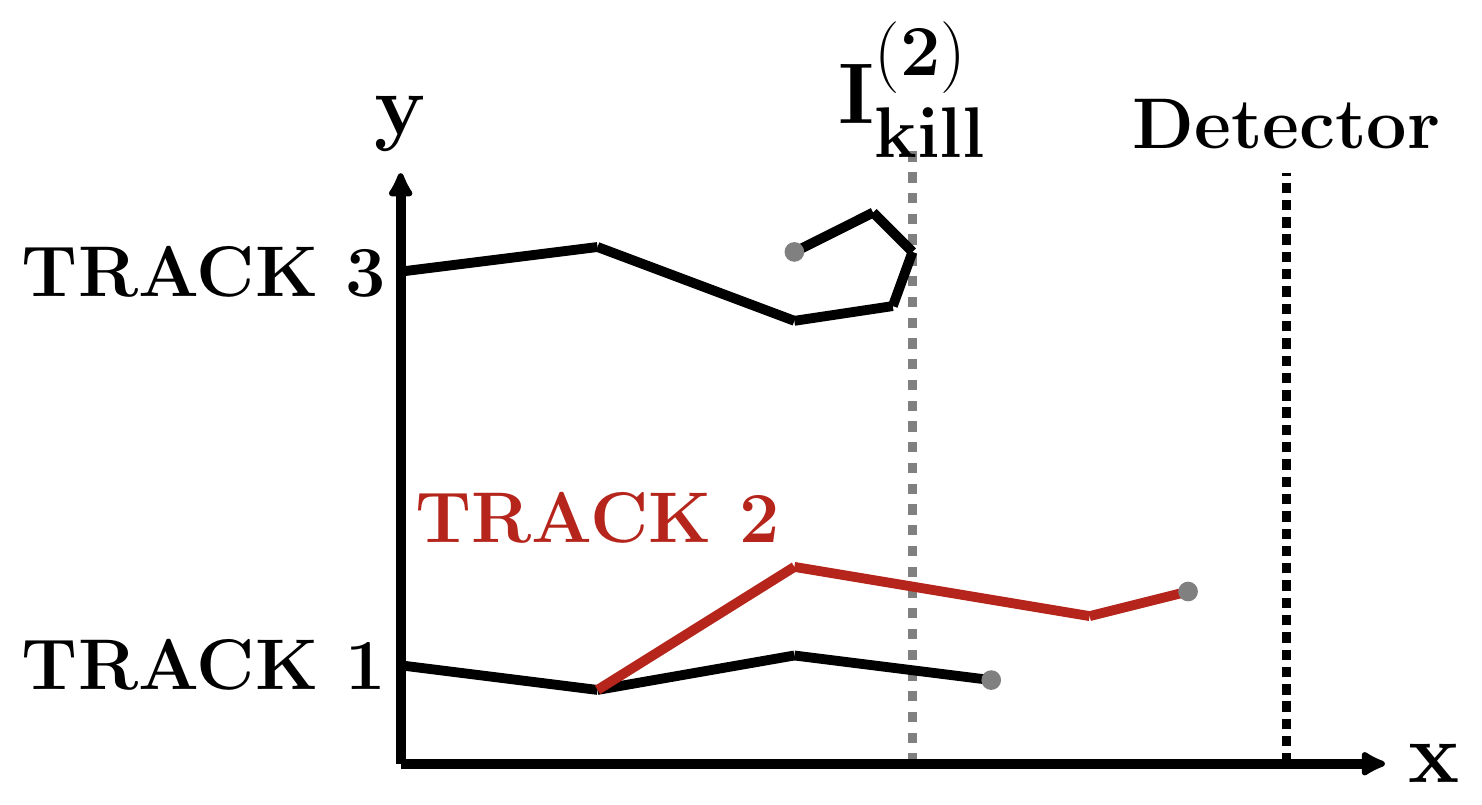}
  \subcaption{Iteration 2}
  \label{fig:AMS_iteration_2}
 \end{subfigure}
 \hfill
 \begin{subfigure}[t]{0.48\textwidth}
  \centering
  \includegraphics[width=1.1\textwidth]{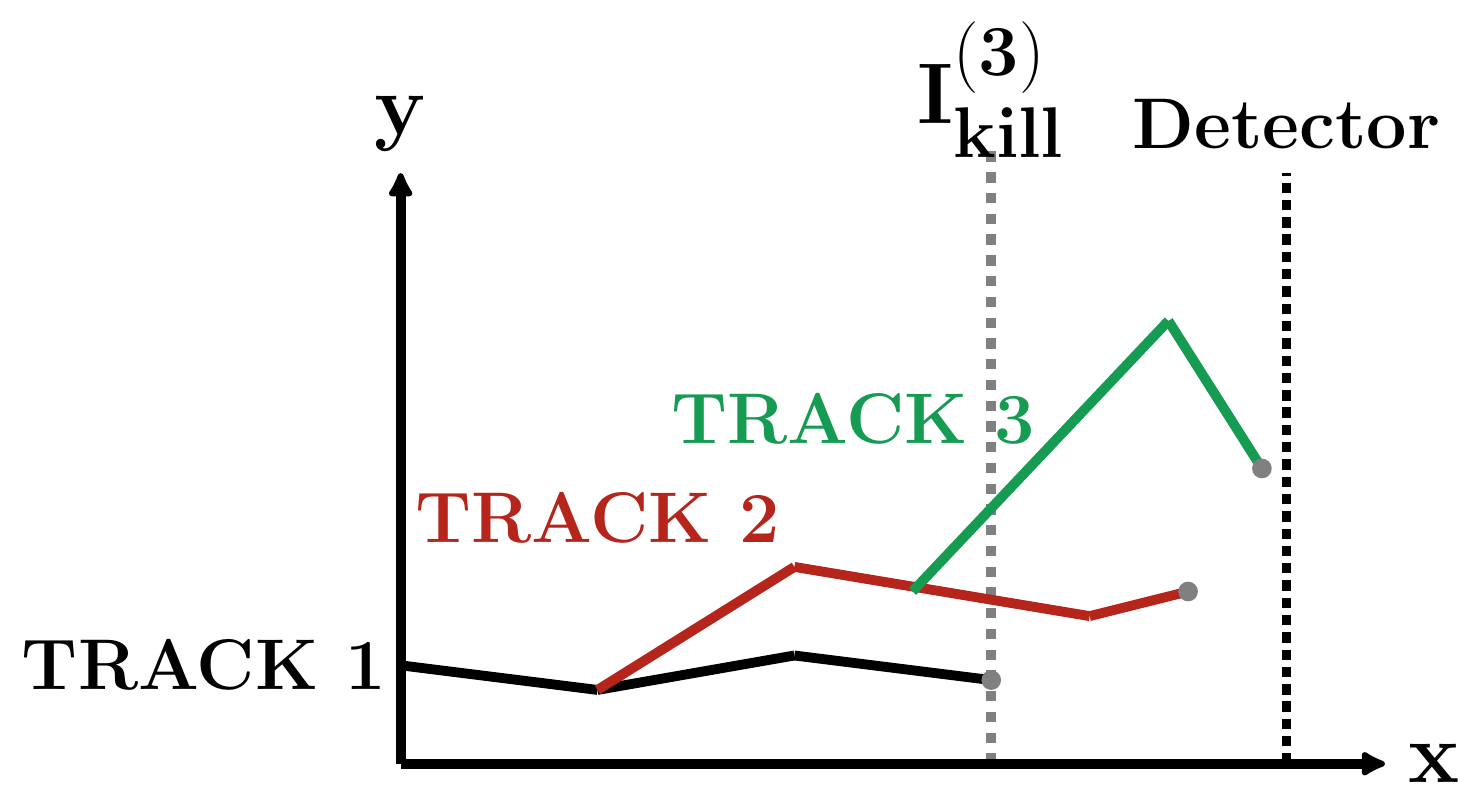}
  \subcaption{Iteration 3}
  \label{fig:AMS_iteration_3}
 \end{subfigure}
 \caption{AMS iterations with a detector defined in the $\left( x,y \right)$ plane, with $N=3$ and $K=1$. The closer the detector, the more important the particle is.}.
 \label{fig:AMS_iterations}
\end{figure*}

To keep the total number of tracks constant, $K_i$ tracks are sampled uniformly among the remaining ones to be duplicated to make up for the deleted ones.
When a track is selected for duplication, the first point of importance greater than the kill level is copied into the new track.
The new track thus created is simulated as described in Section \ref{sec:AMS_transport_step}.
This is illustrated by Figure \ref{fig:AMS_iterations}.

Once the AMS algorithm has stopped (we note the last iteration $I$), scores are computed using the following estimator
\begin{equation}
    \widetilde{\phi}_\textrm{detector} = \phi_\textrm{detector}^{(I)}\times \alpha_{AMS}.
\end{equation}
where $\widetilde{\phi}_\textrm{detector}$ is an unbiased estimator of $\phi_\textrm{detector}$, $\phi_\textrm{detector}^{(I)}$ is the estimation of score $\phi_\textrm{detector}$ based on all iterations tallies using classical Monte Carlo estimators (e.g., track length or collision estimators) and $\alpha_{AMS}$ is defined by
\begin{equation}\label{eq:alpha_AMS}
    \alpha_\textrm{AMS} \equiv \prod_{i=1}^{I} \left( 1 - \frac{K_i}{N} \right).
\end{equation}
While $\alpha_\textrm{AMS}$ is used to correct tallies so that results remain unbiased, it can also be interpreted as an estimation of the probability to reach the detector.
Although it is not described here, it is possible to define an \emph{on-the-fly} scoring procedure to compute scores outside the detector.

\subsubsection{About the importance function}\label{sec:AMS_importance_function}
The importance function is a function that maps $\mathbb{R}^L \rightarrow \mathbb{R}$, where $L$ is the number of parameters considered to compute the importance (position, direction, energy, ...). 
Its purpose is to rank tracks in the AMS (see Section \ref{sec:AMS_ranking_step}) and should be chosen to push neutron histories towards the detector. 
The optimal choice for that function should lead to the best estimation of the probability $\alpha_\textrm{AMS}$ leading to the minimum variance \cite{Brehier2019OnSplitting}. 
In particle transport, although there is no formal demonstration for the AMS, the solution of the adjoint Boltzmann equation for the detector \cite{lewins1965importance} is generally considered to be the optimal choice.

Although there are no further requirements, it is best to avoid importance functions presenting discrete levels that could lead to aggregates of particles.
Indeed, for a discrete importance function, if the $N-K+1$ particles with the highest importance were on the same level, the splitting level would be equal to $I^{(i)}_{MAX} < \infty$, and the iterations would stop before enough particles have reached the detector.

Finally, since the importance function is only used to rank particles, only the relative importance between two particles matters, making the AMS a reasonably robust and easy to use method \cite{louvin2017development}.

\subsection{Adaptation to criticality calculations}\label{sec:AMS_criticality}
For subcritical systems, the neutron population tends to go extinct with time/generations\footnote{One can artificially define a generation as a neutron trajectory between birth and death by absorption or leakage. 
Therefore, neutrons born by fission are considered in the next generation of the particle that caused the fission.}.
Therefore, it is clear that the lower the $k_\textrm{eff}$, the less likely a neutron history is to survive over several generations, and reaching a distant generation is then a rare event.
In criticality calculations, the AMS is used to re-sample histories and push them across generations.
Moreover, it has been specified in Section \ref{sec:AMS_transport_step} that any collision point could be added to the AMS structure, which implies that any collision point could be used for the re-sampling of neutrons.
Compared to the power iteration, where new neutrons are sampled at fission sites only, this induces a non-negligible difference in terms of the precise equation solved by the algorithm due to spectral and spatial heterogeneity effects as explained by Cullen et al. in Ref. \cite{cullen2003static}. 
However, to ease the comparison with the power iteration method, only fission points can be saved, as it is possible to store any stopping point as long as the system remains Markovian.

\begin{figure*}[!htb]
 \centering
 \begin{subfigure}[t]{0.48\textwidth}
  \centering
  \includegraphics[width=1.05\textwidth]{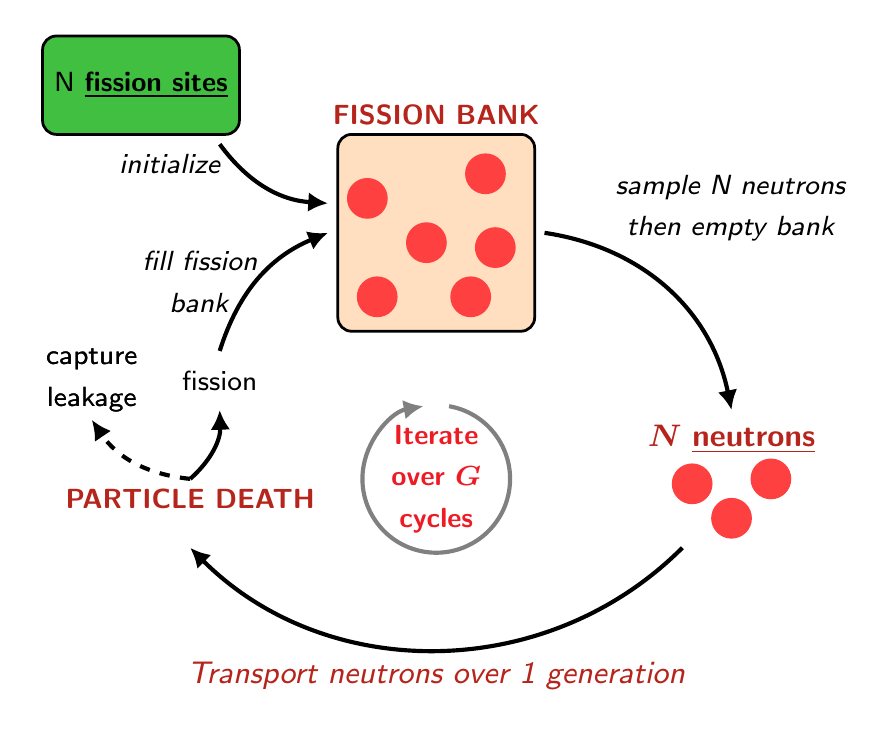}
  \subcaption{PI: iterates over fission neutrons (1 generation per iteration) after being initialized with an arbitrary fission distribution. Neutrons are sampled from fission neutrons of the last iteration.}
  \label{fig:PI}
 \end{subfigure}
 \hfill
  \begin{subfigure}[t]{0.48\textwidth}
  \centering
  \includegraphics[width=1.05\textwidth]{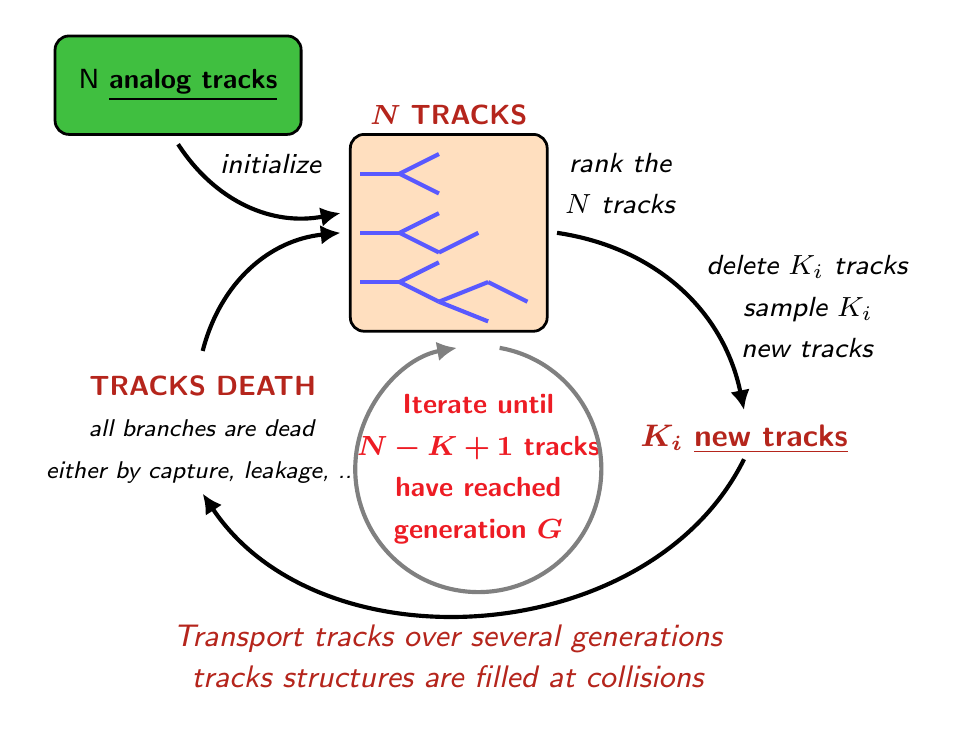}
  \subcaption{AMS in criticality: iterates over re-sampled tracks (multiple generations per iteration) after being initially fed with $N$ analog tracks.}
  \label{fig:AMS}
 \end{subfigure}
 \caption{Comparison scheme between the Power Iteration (PI) and AMS used for criticality.}
 \label{fig:PI_vs_AMS}
\end{figure*}

In a subcritical system, $k_\textrm{eff}$ can be interpreted as the probability for a neutron to go from one generation to the next.
With that in mind, the probability for one neutron history with initially one particle to reach generation $G$ is
\begin{equation}\label{eq:survival_probability}
    P_\textrm{survive}(G) = k_\textrm{eff}^G.
\end{equation}
Hence, tracking neutrons over generations in a subcritical system can also be considered as an attenuation problem over generations when no population control is done, making this a suitable scope for using the AMS. The idea is then to define a detector in generation (i.e., a target generation towards which neutrons will be pushed by the re-sampling algorithm) and to track neutrons not in time but over successive generations.
For this purpose, in the subsequent sections of this article, the importance function used to rank tracks will be of the following form
\begin{equation}\label{eq:generation_importance}
    I(\pmb{r}, g) = g + f(\pmb{r})
\end{equation}
where $g$ is the neutron generation to push neutrons over generations, and $0 \leq f(\pmb{r}) \leq 1$ is a function of space used to discriminate neutrons of the same generation so that the importance function is not discrete (see Section \ref{sec:AMS_importance_function}).
The resulting algorithm is compared to the Power Iteration in Figure \ref{fig:PI_vs_AMS}.

The probability of reaching the detector defined by Equation \ref{eq:alpha_AMS} could therefore be compared to the $k_\textrm{eff}$ given as a probability of survival (see Equation \ref{eq:survival_probability})
\begin{equation}
    \alpha_\textrm{AMS} = \prod_{i=1}^{I} \left( 1 - \frac{K_i}{N} \right) = k_\textrm{eff}^G
\end{equation}
where $I$ is the total number of AMS iterations, and $G$ is the target generation.
Therefore we can build the following estimation of the $k_\textrm{eff}$
\begin{equation}
    k_\textrm{eff} = \left\{ \prod_{i=1}^{I} \left( 1 - \frac{K_i}{N} \right) \right\}^{1/G}.
\end{equation}

While this holds for subcritical systems, it is no longer valid when the system is critical or supercritical. 
In the description above, the neutron population is not constrained by some population control mechanism, which allows for fluctuations in the system's number of particles. 
This mechanism can induce population growth for systems close to criticality (and supercritical systems), thus increasing the number of branches inside a track. 
Since it is necessary to take those branches into account during the re-sampling step, the number of re-sampled branches may increase as iterations go by, leading to slower and slower iterations (as well as more and more memory used).
Besides, as in fixed source calculations, neutron histories must end at some point to be able to rank the tracks as previously described, which could be troublesome in critical and supercritical regimes.
Thus, the branchless collision method \cite{lux1991monte} was used to limit the number of branches inside a track (equal to one if no splitting is used), preventing these issues. 
For a system with leakage, making particles carry population fluctuations through statistical weights produces a numerically subcritical system (no particles are produced by splitting, and some disappear by leaking out of the geometry). 
Hence, it is possible to reproduce a population attenuation over generations that appeals to the use of the AMS whatever the $k_\textrm{eff}$ if the branchless collision method is used.

\FloatBarrier
\section{Application to a one-dimensional slab reactor}\label{sec:results}
To characterize the AMS behavior regarding criticality calculations, the method was tested on a one-dimensional bare slab reactor.
We tested the method on a simple case in one dimension with mono-energetic neutrons, thus limiting the number of particles needed to explore the space and allowing us to compare results to a simple analytical solution.
\subsection{Bare slab properties}
The modeled system is a one dimensional homogeneous bare slab reactor with leakage on the sides, the total size of the slab being 100 cm, from $x_{min} = -50.0\textrm{ cm}$ to $x_{max} = 50.0\textrm{ cm}$.
The slab size was chosen so the system would be loosely coupled considering the cross sections of the system, presented in Table \ref{tab:XS_1D_rod}.
Three reactions are possible following a collision in analog transport: fission, capture, and isotropic scattering.
The cross sections were arbitrarily chosen to model a slightly supercritical system to assess the capability of the AMS to model supercritical systems using the branchless collision method, the resulting $k_\textrm{eff}$ being equal to $1.03437$.
\begin{table}[!htbp]
  \centering
\caption{Physical properties for homogeneous 1D rods}
    \label{tab:XS_1D_rod}
    \begin{tabular}{l|c}
        \toprule
        Mean number of fission neutrons ($\Bar{\nu}$) & $2.383$  \\
        Neutron speed ($v$) & $2.2\times10^{4}\ cm.s^{-1}$ \\
        \midrule
        \multicolumn{2}{c}{Macroscopic cross sections} \\
        Fission ($\Sigma_f$) & $0.250\ cm^{-1}$ \\
        Absorption ($\Sigma_a$) & $0.575\ cm^{-1}$ \\
        Scattering ($\Sigma_s$) & $0.425\ cm^{-1}$ \\
        Total ($\Sigma_{tot}$) & $1.00\ cm^{-1}$ \\
        \bottomrule
    \end{tabular}
\end{table}
The simplicity of the system also allowed us to compute an analytical solution using diffusion theory \cite{zweifel1977reactor}, which is used as a comparison in the following results
\begin{equation}
    \phi(x) = \phi_0 \cos\left( \frac{\pi}{2(a+z_0)}x \right)
\end{equation}
with $\phi_0$ depending on the normalization, $a$ being the reactor half size (here $a = 50.0\ cm$) and $z_0$ is the linear extrapolated end point of the reactor and is defined as
\begin{equation}
    z_0 = \frac{2}{3\Sigma_{tr}}
\end{equation}
where $\Sigma_{tr}$ is the transport cross section, which is equal to the total macroscopic cross section $\Sigma_t$ since all collisions are isotropic in the laboratory referential.

Different calculation options were tested and compared to assess the effects of each regarding clustering and variance estimation.
The sets of options corresponding to each case are described in Table \ref{tab:calculations}.
Four different simulations were done for the power iteration to distinguish between the effects of the methods used.
As a matter of fact, collisions were simulated either by in an analog way or by using the branchless collision method.
As for the population control operated between cycles, two sampling methods were used, a simple sampling with replacement and the combing method \cite{booth1996weight}.
For the AMS, since the system is supercritical, no simulation with analog collisions was done since it would lead to a divergence of the particle number over generations, hence computation cost issues.

\begin{table}[!htbp]
  \centering
\caption{Description of calculation parameters}
    \label{tab:calculations}
    \begin{tabular}{l c c c}
        \toprule
        Case & Population control & Collisions & Importance (AMS only)  \\
        \midrule
        PI analog & sampling with replacement & analog & \\
        PI branchless & sampling with replacement & branchless & \\
        PI combing & combing & analog & \\
        PI combing branchless & combing & branchless & \\
        AMS branchless & AMS & branchless & $g + \cos\left( \frac{\pi x}{2a}\right)$ \\
        \bottomrule
    \end{tabular}
\end{table}

All the calculations presented below started from a uniform fission distribution and were done with 1000 neutrons per cycle ($N = 1000$ initial independent tracks for the AMS) over $G = 1000$ successive generations in $M = 1000$ independent runs.
Having too few particles per generation to facilitate clustering in all cases was deliberate to study the effects of methods on neutron clustering.

Finally, estimators used in this work for the flux and the $k_\textrm{eff}$ rely on the \emph{on-the-fly} scoring procedure detailed in Ref. \cite{louvin2017development}. 
In each generation $i$, the flux was computed using the collision estimator and normalized so its spatial shape could be averaged over successive generations.
The $k\textrm{eff}$ estimator is based on the physical interpretation of the $k_\textrm{eff}$, and was computed as the ratio of neutrons produced in a generation over the ones produces in the previous generation.

\subsection{Convergence of inactive cycles and behavior of the Shannon entropy}
\begin{figure*}[!htb]
 \centering
  \includegraphics{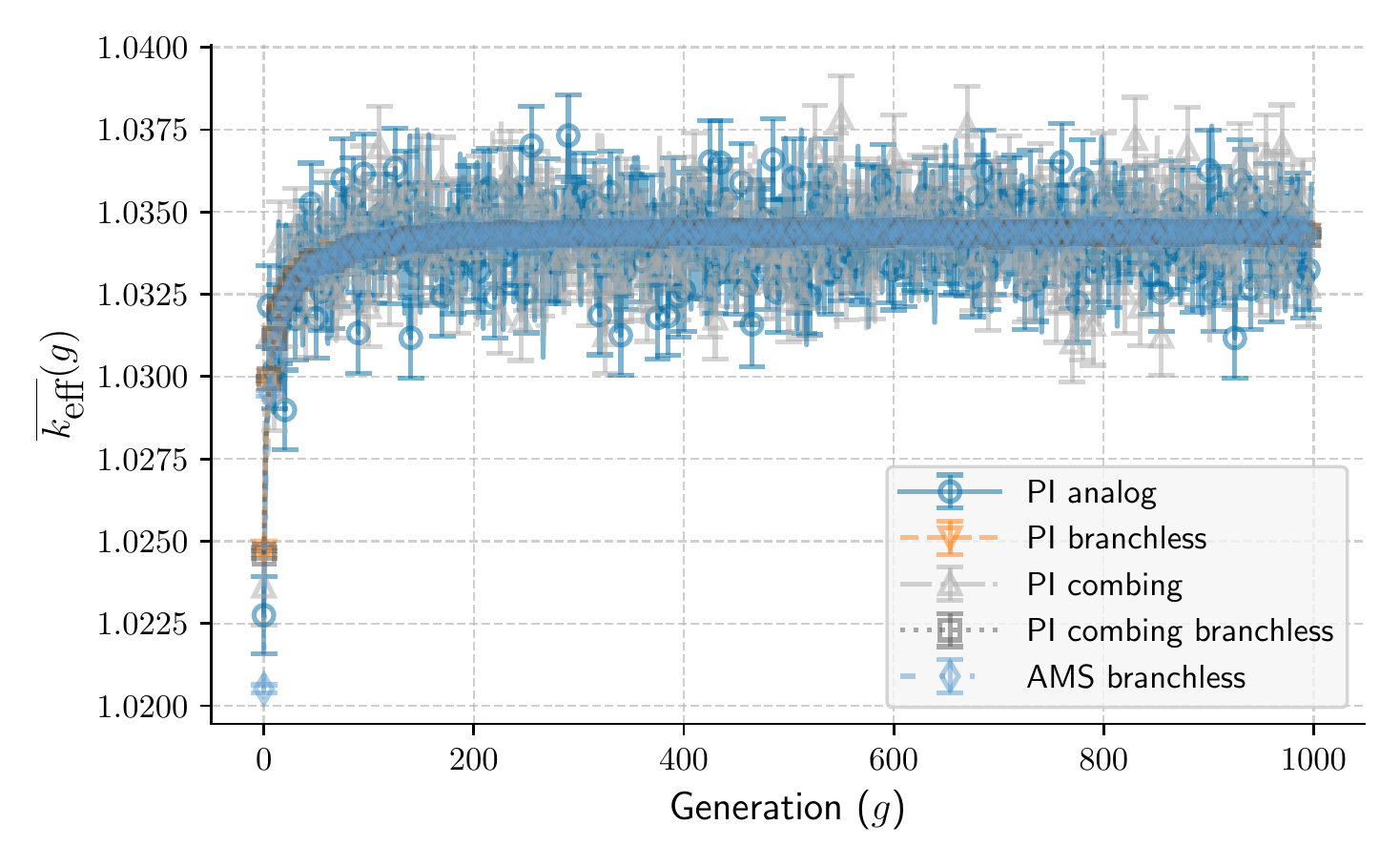}
 \caption{Convergence of the average $k_\textrm{eff}$ (over 1000 independent runs) with $3\sigma$ confidence intervals. Cases PI branchless, PI combing branchless, and AMS branchless show the same results, with narrow confidence intervals.}
 \label{fig:keff_convergence}
\end{figure*}

Firstly, the convergence of the $k_\textrm{eff}$ estimates in each generation, and the Shannon entropy \cite{brown2006use} of the system were considered to set the number of inactive cycles. 
Firstly, the convergence of the $k_\textrm{eff}$ is shown on Figure \ref{fig:keff_convergence} as the mean $k_\textrm{eff}$ per generation, computed in $M$ independent simulations, as a function of the cycle number
\begin{equation}
    \overline{k_\textrm{eff}}(g) = \frac{1}{M}\sum_{m=1}^{M} \frac{N_g^{(m)}}{N_{g-1}^{(m)}}
\end{equation}
where $N_g^{(m)}$ is the number of neutrons born in generation $g$ for simulation $m$.
Its convergence is quite fast for every calculation. Apart from statistical fluctuations that have no impact on the mean value, as seen later, all methods seem to converge towards the same value.

\begin{figure*}[!htb]
 \centering
  \includegraphics{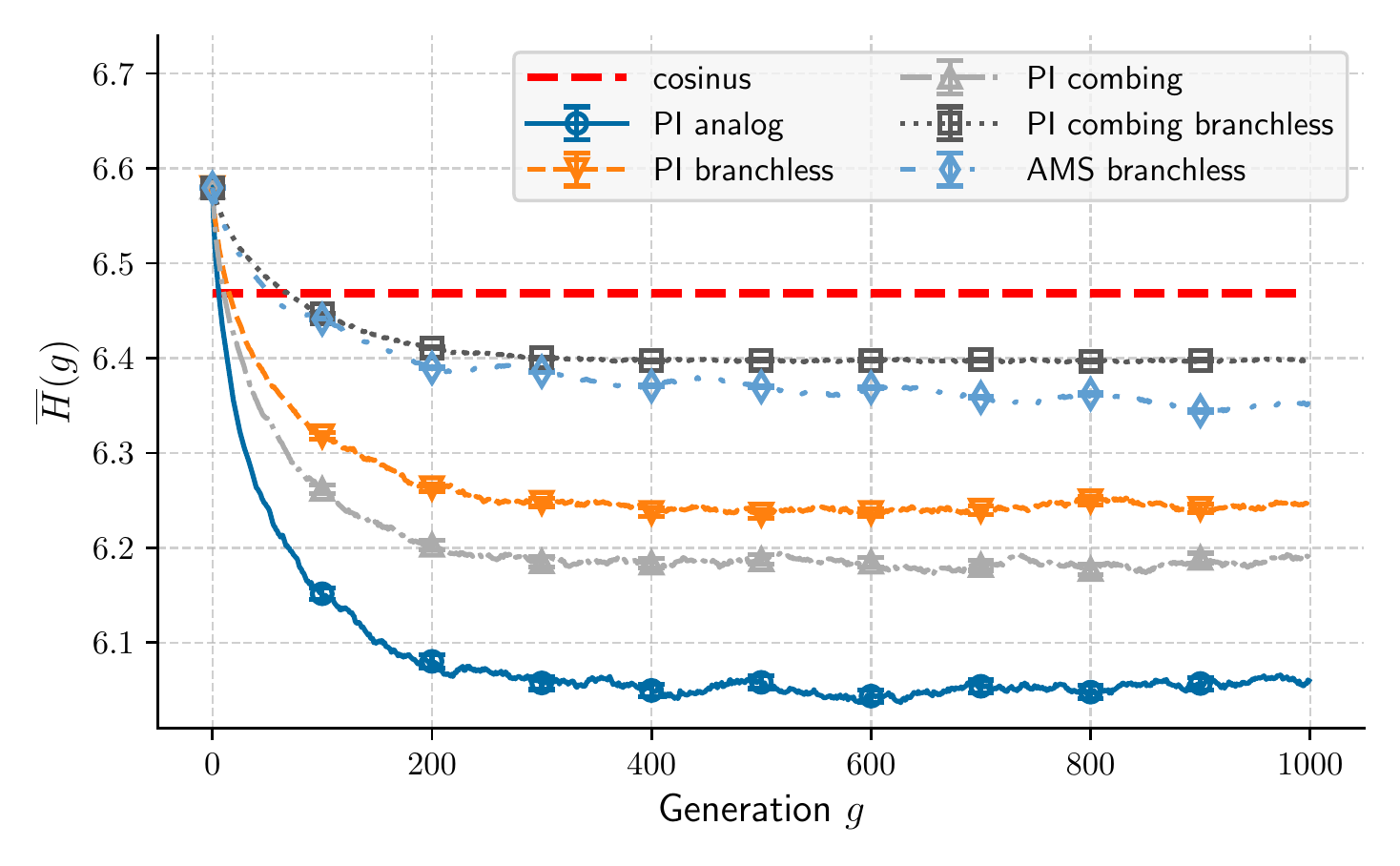}
 \caption{Evolution of the mean entropy over generations.}
 \label{fig:entropy_convergence}
\end{figure*}

Secondly, the Shannon entropy was used to assess the spatial flux convergence \cite{brown2006use} and set the number of generations that have been discarded when computing average scores.
Its averaged value over $M$ independent simulation has been computed in each generation as such
\begin{equation}\label{eq:entropy}
    \overline{H}(g) = \frac{1}{M} \sum_{m=1}^M
    \left[ -\sum_{l=1}^{N_\textrm{bins}}
    \frac{\phi_g^{(m)}(x_l)}{\phi_{g, tot}^{(m)}}
    \log_2\left( \frac{\phi_g^{(m)}(x_l)}{\phi_{g, tot}^{(m)}} \right)
    \right]
\end{equation}
where $N_\textrm{bins}$ is the number of spatial bins along the $x$ (here 100 bins), $\phi_g^{(m)}(x_l)$ is the normalized flux estimated in generation $g$ in bin $x_l$ for simulation $m$, and $\phi_{g,tot}^{(m)}$ is the normalized flux at cycle $g$ for simulation $m$ integrated over $x$.
The results are plotted in Figure \ref{fig:entropy_convergence}.
As expected, the entropy converges slower than the $k_\textrm{eff}$.
All cases were considered to have reached an acceptable convergence for $g=200$.
For the rest of the article, the number of inactive cycles was set to 200.

\begin{figure*}[!htb]
 \centering
  \includegraphics[width=1\textwidth]{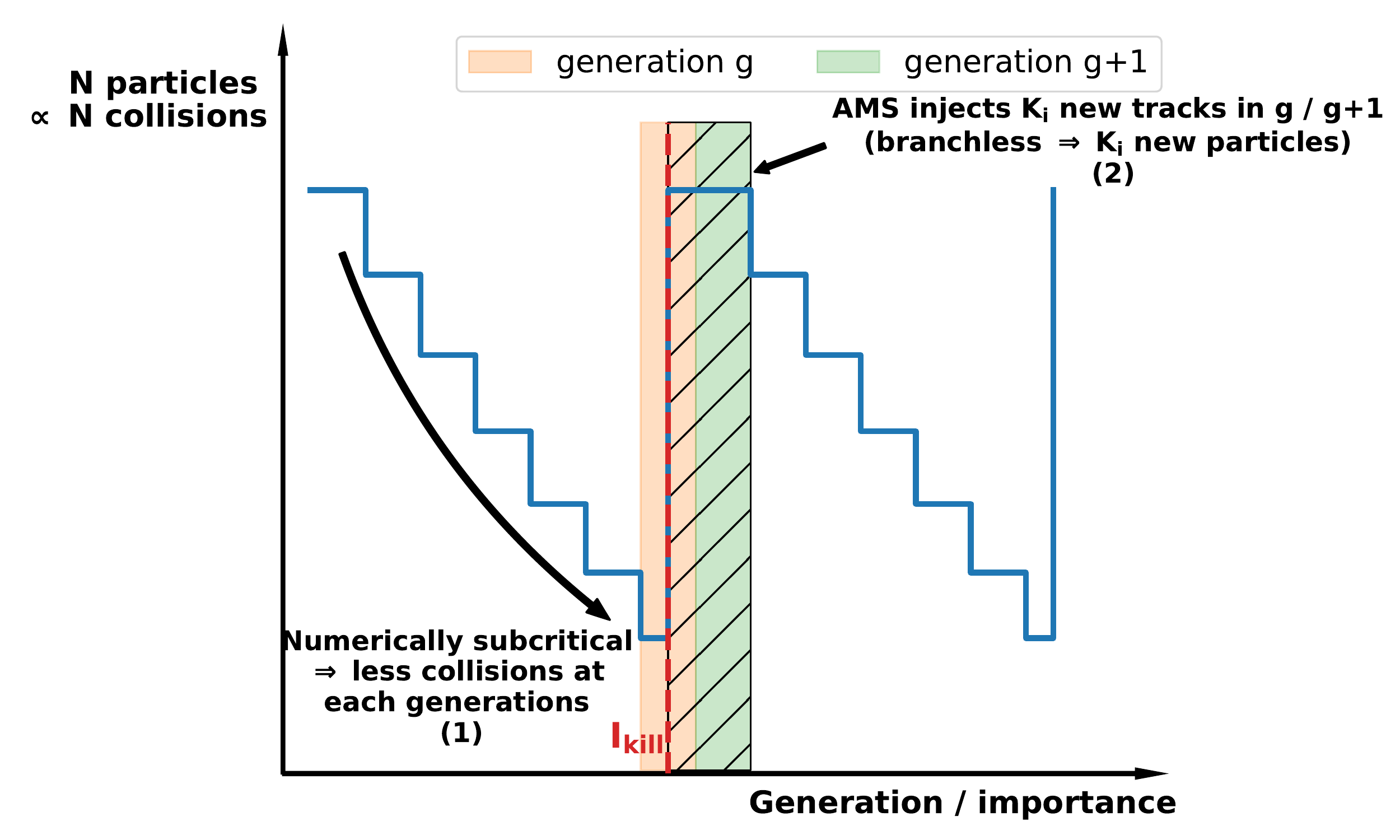}
 \caption{AMS population control mechanisms. In this example, all neutrons that were re-sampled appeared in generation $g$.}
 \label{fig:AMS_pop_ctrl}
\end{figure*}

\begin{figure*}[!htb]
 \centering
  \includegraphics{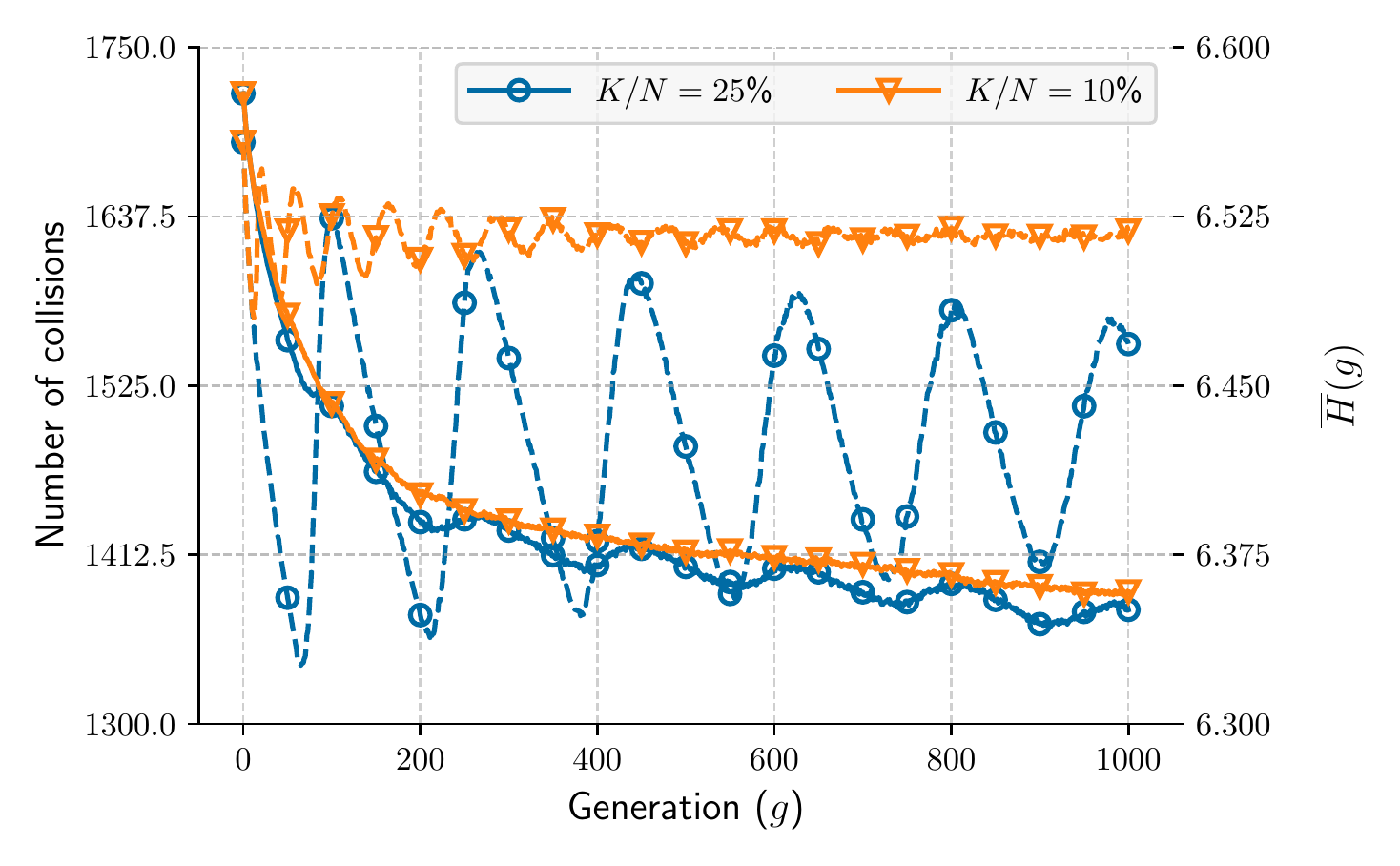}
 \caption{Mean entropy (solid lines) and mean number of collision points (dashed lines) per generation for the AMS + branchless case, for $K/N = 10\%$ and $K/N = 25\%$.}
 \label{fig:entropy_AMS}
\end{figure*}

Unlike the $k_\textrm{eff}$, the Shannon entropy, as defined in Equation \ref{eq:entropy}, and presented in Figure \ref{fig:entropy_convergence}, does not converge to the same asymptotic value for all the different methods and is lower than the theoretical value for a cosine shape distribution in all cases.
Moreover, the entropy presents oscillations when the AMS is used. 
This phenomenon is likely due to how the AMS injects particles into the simulation and can be decomposed into two underlying mechanisms: the numerical subcriticality of our system (1) and the re-sampling of new particles by the AMS (2), as portrayed in Figure \ref{fig:AMS_pop_ctrl}.
Indeed, as explained in Section \ref{sec:AMS_criticality}, the system was set to be numerically subcritical thanks to the branchless collision method to model an attenuation problem for the AMS.
Thus, without population control, the population progressively goes extinct over generations (see mechanism (1) on Figure \ref{fig:AMS_pop_ctrl}).
As for the re-sampling of new particles, detailed in Section \ref{sec:AMS_sampling_step}, the algorithm samples about $K$ new tracks close to an importance level defined by the kill level of the current iteration. 
Since the generation of a particle mainly drives the importance value, as stated by Equation \ref{eq:generation_importance}, we have
\begin{equation}
    g \leq I_\textrm{kill}(i) \leq g+1.
\end{equation}
The new tracks sampled by the algorithm will therefore start either in generation $g$ with an importance higher than $I_\textrm{kill}(i)$, or in generation $g+1$ (illustrated by the dashed area on Figure \ref{fig:AMS_pop_ctrl}).
This will result in a much higher increase in the total population sampled in these generations (see mechanism (2) on Figure \ref{fig:AMS_pop_ctrl}). 
Intuitively, the more particles in the system, the closer (and smoother) their distribution will be to the natural distribution.
Hence, as the system loses particles, the flux estimation gets noisier due to statistical fluctuations.
Although these fluctuations have no visible impact on the average estimate, they slightly modify the entropy.
Consequently, when the number of particles in the system is increased by the re-sampling of the AMS in generations $g$ and $g+1$, the flux estimate in generation $g+1$ gets smoother than in previous generations, inducing an increase in the entropy (which will then decrease as particles disappear, until the next re-sampling step, and so on).
To reduce the amplitude of these fluctuations, one could reduce the number of re-sampled particles at each iteration by reducing the value of $K$.
It would also increase the frequency of the entropy since the re-sampling of particles would occur more frequently.
To illustrate the phenomenon, the mean number of collisions (unweighted) and the corresponding entropy per generation were computed and plotted on Figure \ref{fig:entropy_AMS} for $K/N = 25\%$ and $K/N = 10\%$.
Making the system less subcritical (numerically) should also reduce the frequency of the oscillations; however, it should not modify the amplitude of the oscillations.
Another way to make the oscillations disappear without changing the simulation would be to compute the Shannon entropy over a coarser spatial mesh, which would lessen the spatial fluctuations.

In a nutshell, the AMS does not seem to lengthen nor abridge the convergence period.
Besides, the observed oscillations of the flux entropy it might produce are natural and do not affect the average flux estimation.
\subsection{Fundamental mode estimates}
The fundamental mode of a multiplicative system described by the $k$-eigenvalue equation is characterized by the highest eigenvalue $k_0 = k_\textrm{eff}$ and the associated eigenvector : the fundamental flux distribution.

\begin{figure*}[!htb]
 \centering
  \includegraphics{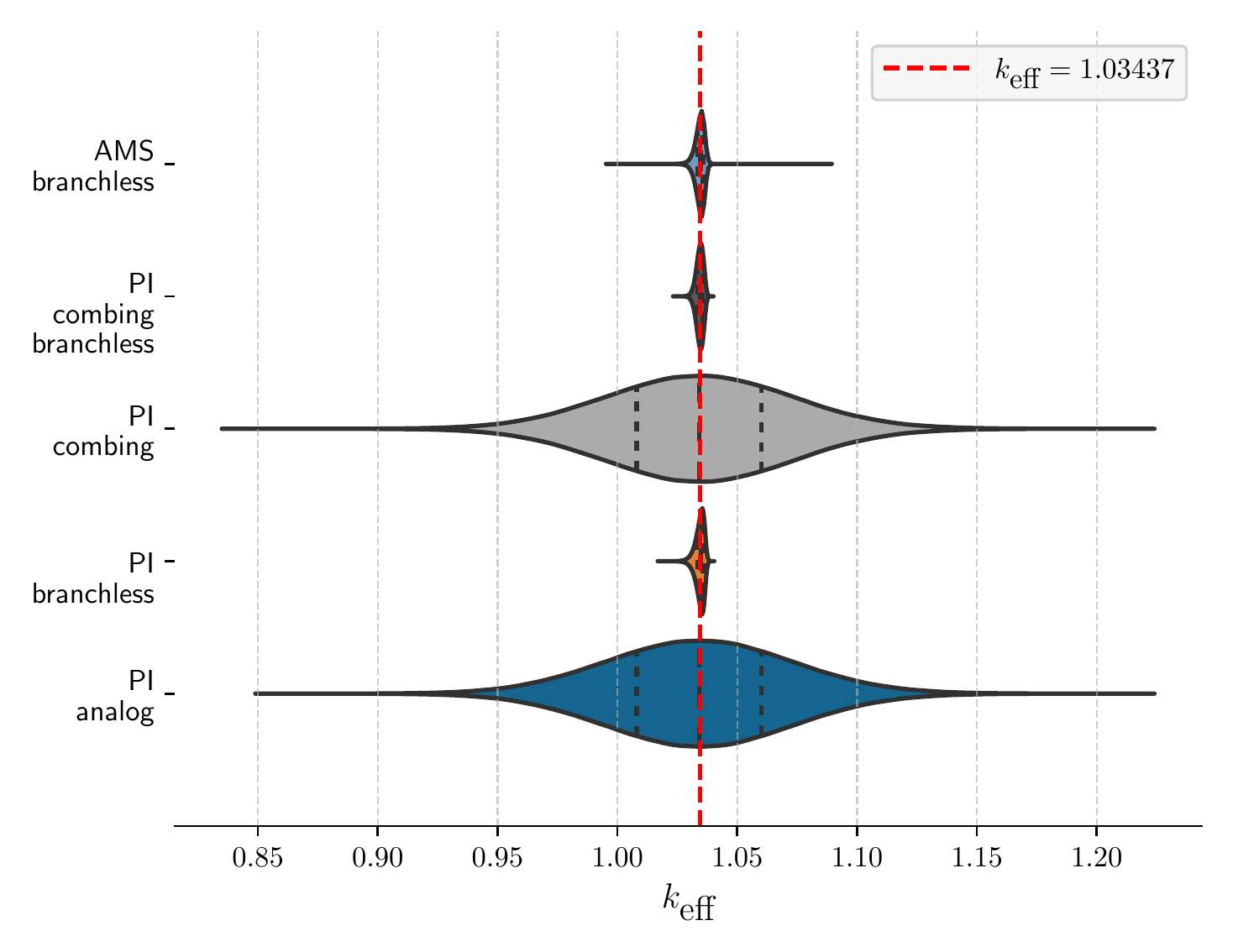}
 \caption{Distribution of the $k_\textrm{eff}$ after convergence. Dashed lines represent the first, second and third quartiles of the empirical distribution.}
 \label{fig:keff}
\end{figure*}

For the system described earlier, the $k_\textrm{eff}$ distribution over 800 actives cycles in 1000 independent simulations is plotted in Figure \ref{fig:keff}.
Cases with branchless collision show quite similar distributions, with much less dispersion around their mean value than for the non-branchless cases.
Hence, regarding the $k_\textrm{eff}$ estimation, the branchless collision method seems to be the main contributor to the variance reduction, while the differences between population control methods are not very significant. 

\begin{figure*}[!htb]
 \centering
  \includegraphics{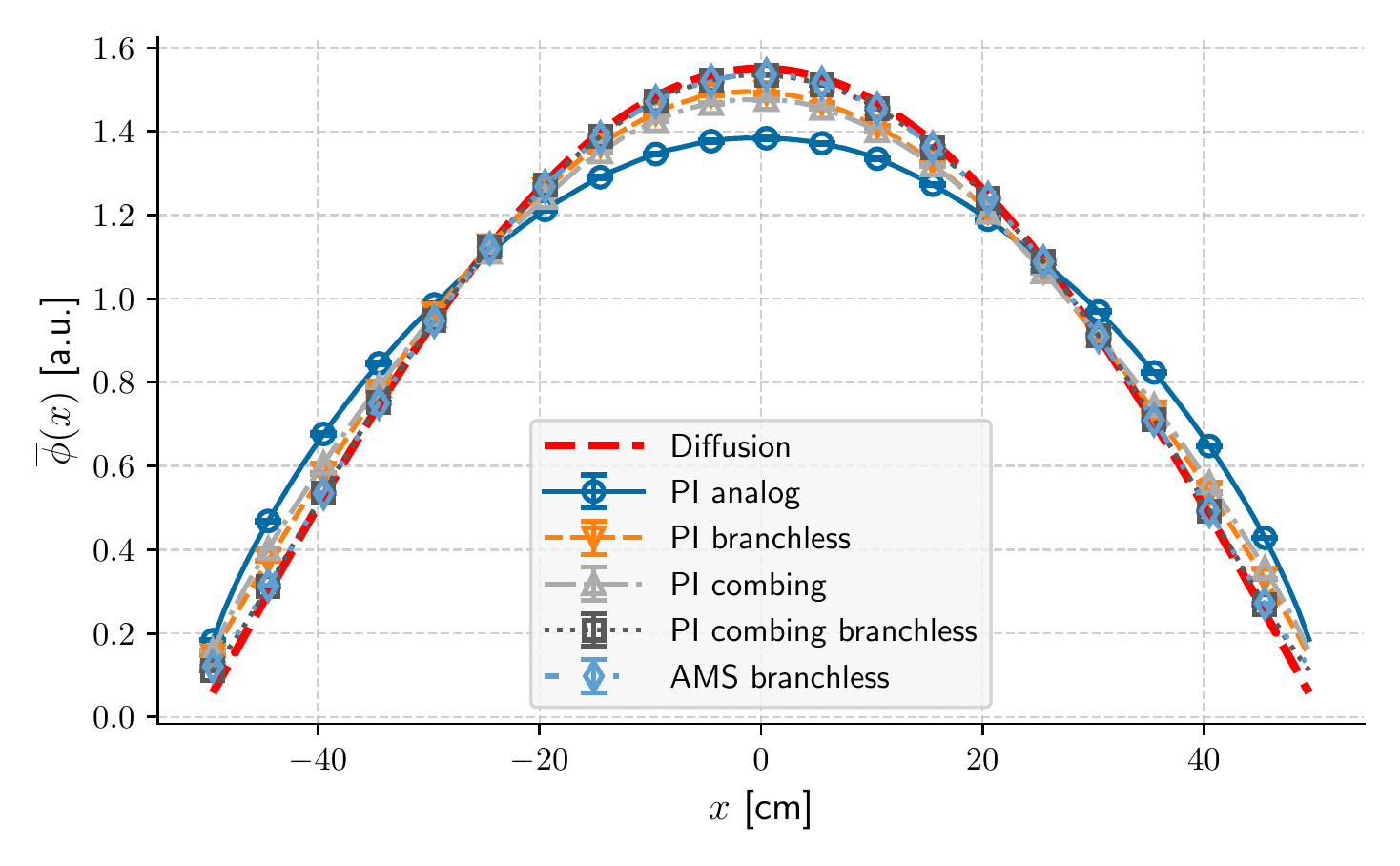}
 \caption{Spatial flux profile with $3\sigma$ confidence interval (top) with the relative (center) and absolute (bottom) discrepancies against the analytical cosine-shaped solution. }
 \label{fig:flux}
\end{figure*}

Besides the eigenvalue, the fundamental flux was also computed over 800 successive generations in 1000 independent simulations, and is plotted on Figure \ref{fig:flux}.
The first striking observation is the lack of consistency between the solutions of the different methods, even with $3\sigma$ confidence intervals. Although they do not include the analytical solution within their $3\sigma$ confidence interval, the cases that show the less difference with the analytical cosine are the power iteration with branchless collision and combing used for population control case, and the AMS combined with branchless case\footnote{The analytical solution was computed using the diffusion theory, this is why minor discrepancies are expected, especially on the sides.}. 

These observed deformations of the flux shape are likely due to clustering effects that affect the estimation of the mean spatial flux.
Since our goal was to compare methods behavior regarding clustering, those effects were expected due to the low number of particles simulated in each batch, and increasing their number would tend to mitigate clustering effects until they disappear.

By flattening the average flux distribution, the presence of neutron clusters should increase the entropy of the average distribution since the entropy of a uniform distribution is higher than the entropy of a cosine distribution.
At the same time, the estimation of the spatial flux in a generation is noisier than its average over multiple generations, which tend to lower the entropy, and clusters in a generation should lower the entropy even more.
This could explain the different asymptotic values displayed in Figure \ref{fig:entropy_convergence}.
Table \ref{tab:entropy} displays the values of the average spatial flux entropy, as well as the asymptotic value to wich the generational entropy converges (the one shown in Figure \ref{fig:entropy_convergence}).
We can see that the the averaged flux entropy is systematically higher than the reference solution entropy due to flatter spatial shapes than the analytical diffusion solution, whereas the asymptotic entropy reached in a generation is systematically lower due to statistical noise (which could be reduced if the number of bins used to compute the entropy were to be decreased).
In that regard, the AMS branchless and PI combing branchless cases present the lowest discrepancies (about the same order of magnitude for both methods) with the analytical solution, which implies less noise in the spatial estimation of the flux and an average value closer to the analytical solution than the other cases.

\begin{table}[!htbp]
  \centering
\caption{Values for the asymptotic entropy reached during source convergence and for the flux distribution averaged over active cycles for each case. The differences are computed with respect to the analytical solution; a negative difference means that the distribution is more ordered than the analytical one (in terms of entropy), while a positive difference means that the distribution is less ordered than the analytical one (closer to a uniform distribution).}
    \label{tab:entropy}
    \begin{tabular}{l c c c c }
        \toprule
        \multirow{2}{*}{Case} & \multicolumn{2}{c}{Converged entropy} & \multicolumn{2}{c}{Averaged flux entropy} \\
        & value & difference & value & difference \\
        \midrule
        Analytical solution & $6.4673$ & - & $6.4673$ & - \\
        PI analog & $6.0525$ & $-4.148 \times 10^{-1}$ & $6.5424$ & $7.511 \times 10^{-2}$ \\
        PI branchless & $6.2427$ & $-2.246 \times 10^{-1}$ & $6.4975$ & $3.020\times 10^{-2}$ \\
        PI combing & $6.1838$ & $-2.835\times 10^{-1}$ & $6.5088$ & $4.154\times 10^{-2}$  \\
        PI combing branchless & $6.3976$ & $-6.969\times 10^{-2}$ & $6.4687$ & $1.375\times 10^{-3}$ \\
        AMS branchless & $6.3629$ & $-1.044\times 10^{-1}$ & $6.4704$ & $3.168\times 10^{-3}$ \\
        \bottomrule
    \end{tabular}
\end{table}

The observed bias on the average flux shape related to the clustering phenomenon is further examined in the following section.
\subsection{Effects of the AMS on clustering}\label{sec:clustering}
\begin{figure*}[!htb]
 \centering
 \begin{subfigure}[t]{0.48\textwidth}
  \centering
  \includegraphics[width=1.1\textwidth]{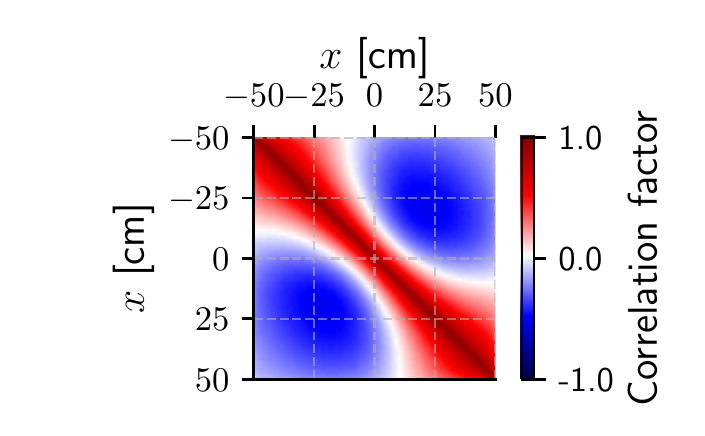}
  \subcaption{PI analog}
  \label{fig:spatial_correl_PI_analog}
 \end{subfigure}
 \hfill
  \begin{subfigure}[t]{0.48\textwidth}
  \centering
  \includegraphics[width=1.1\textwidth]{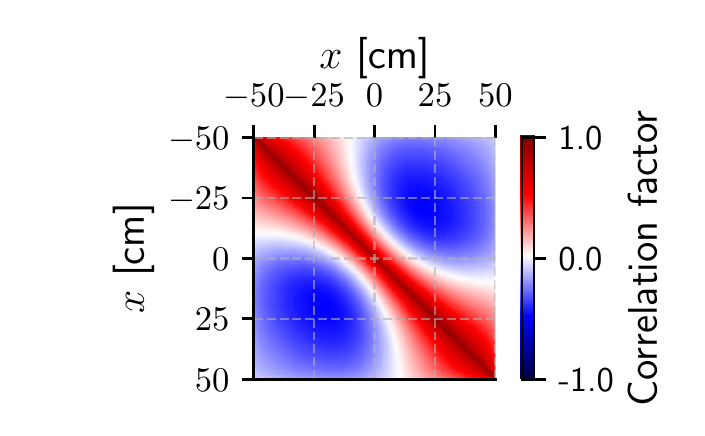}
  \subcaption{PI combing}
  \label{fig:spatial_correl_PI_combing}
\end{subfigure}
\hfill
  \begin{subfigure}[t]{0.48\textwidth}
  \centering
  \includegraphics[width=1.1\textwidth]{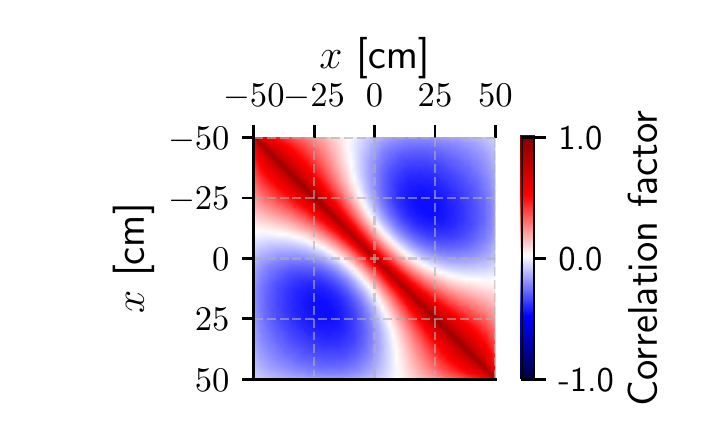}
  \subcaption{PI branchless}
  \label{fig:spatial_correl_PI_branchless}
\end{subfigure}
 \caption{Spatial correlations (scale set from $-1$ to $1$)}
 \label{fig:spatial_correlations}
\end{figure*}

\begin{figure*}[!htb]
 \centering
  \begin{subfigure}[t]{0.48\textwidth}
  \centering
  \includegraphics[width=1.1\textwidth]{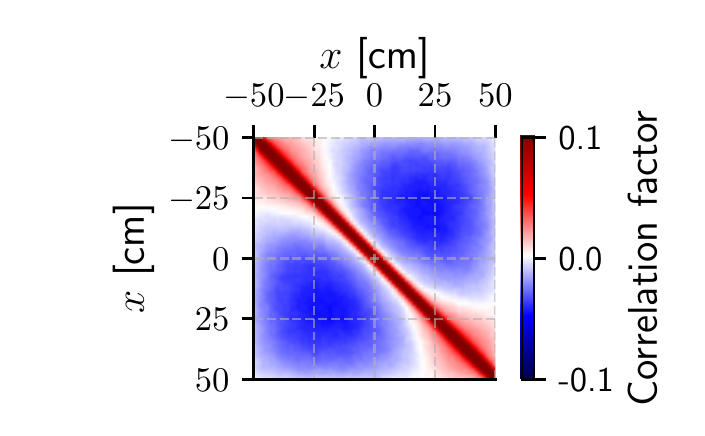}
  \subcaption{PI combing branchless}
  \label{fig:spatial_correl_PI_combing_branchless_vmin01}
  \end{subfigure}
\hfill
  \begin{subfigure}[t]{0.48\textwidth}
  \centering
  \includegraphics[width=1.1\textwidth]{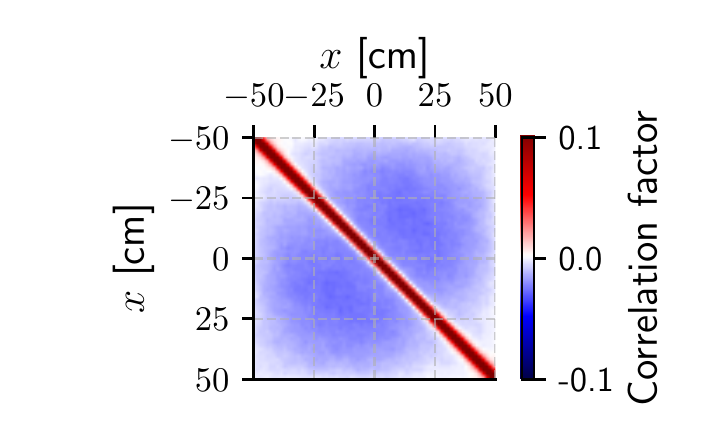}
  \subcaption{AMS branchless}
  \label{fig:spatial_correl_AMS_branchless_vmin01}
\end{subfigure}
 \caption{Spatial correlations (scale set from $-0.1$ to $0.1$)}
 \label{fig:spatial_correlations_vmin01}
\end{figure*}

To assess the probability for clusters to appear, spatial correlations were computed using empirical estimations of Pearson's correlation coefficient between each spatial bin defined by
\begin{equation}
    \rho_{ij} = \frac{Cov\left[ \phi(x_i), \phi(x_j) \right]}{\sigma\left[\phi(x_i)\right] \sigma\left[\phi(x_j)\right]}
\end{equation}
where $Cov\left[ \phi(x_i), \phi(x_j) \right]$ is the covariance between flux estimations in spatial bins $x_i$ and $x_j$, and $\sigma\left[\phi(x_i)\right]$ is the standard deviation of the flux in spatial bin $x_i$.
The results are presented on Figures \ref{fig:spatial_correlations} and \ref{fig:spatial_correlations_vmin01} for 100 spatial bins.
While cases PI analog, PI branchless, and PI combing show similar levels of spatial correlations (see Figure \ref{fig:spatial_correlations}), combing branchless and AMS branchless cases present almost nonexistent spatial correlations as seen in Figure \ref{fig:spatial_correlations_vmin01}.
Since these correlations are deeply linked to the probability for clusters to form \cite{mulatier2015random}, this implies that the two above mentioned cases are the least likely to present neutron cluster problems.
High correlation levels are linked to the number of correlated pairs of particles in the system, whose number increases as generations go by because of independent family\footnote{A neutron family is defined as the set of all neutrons descending from the same ancestor amongst neutrons initially present.} extinctions \cite{dumonteil2017clustering,Sutton2022TowardClustering}.

\begin{figure*}[!htb]
 \centering
  \includegraphics{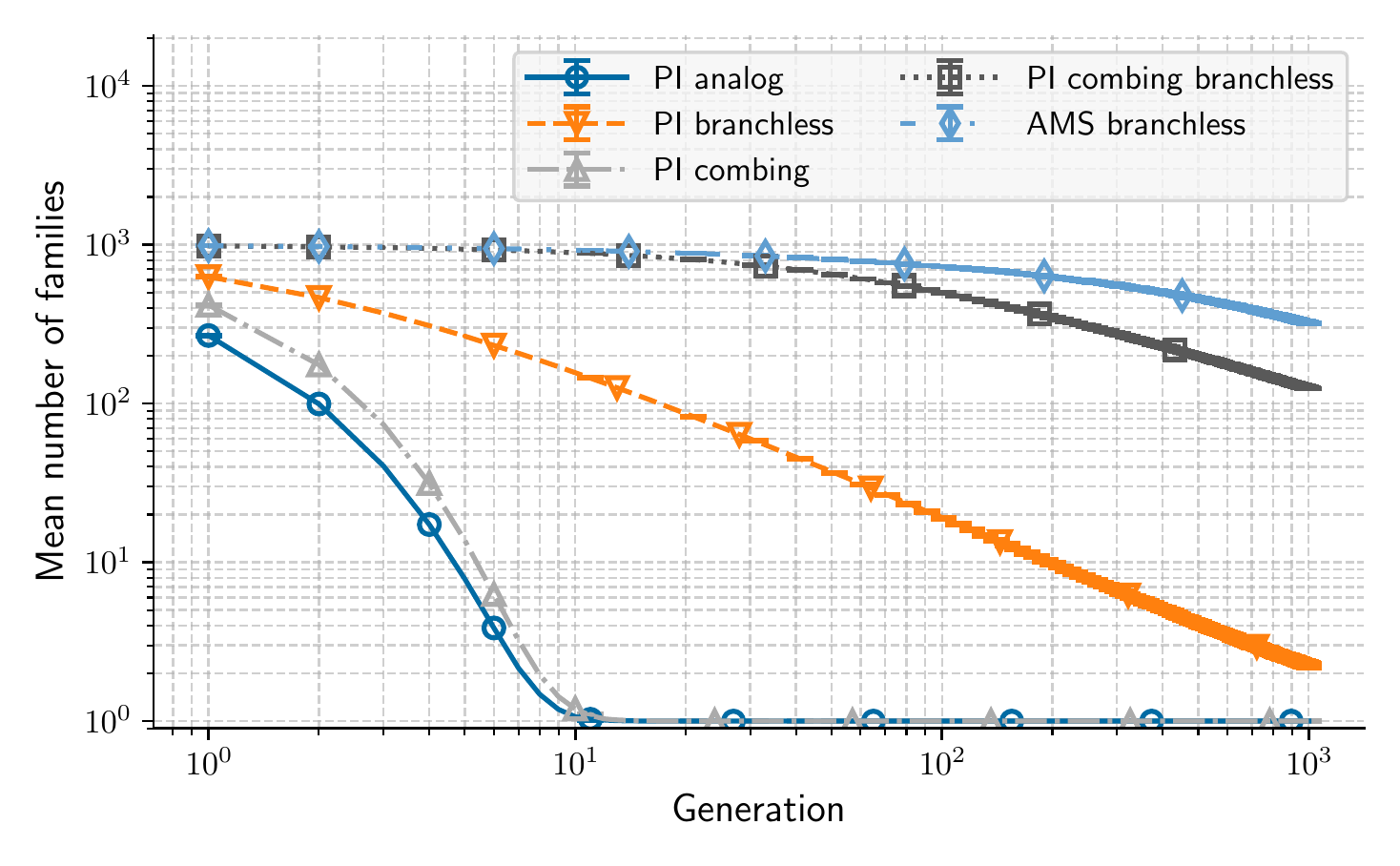}
 \caption{Mean number of families over generations (the number of families at generation 0 is equal to 1000 for all cases).}
 \label{fig:families}
\end{figure*}

Regarding the loss of independent neutron families, Figure \ref{fig:families} shows that both the branchless collision method and population control play a nonnegligible role in preserving uncorrelated pairs of particles.
Indeed, both the combing method and the AMS seem to allow for more neutron lineages to be conserved over generations.

\begin{figure*}[!htb]
  \centering
  \includegraphics{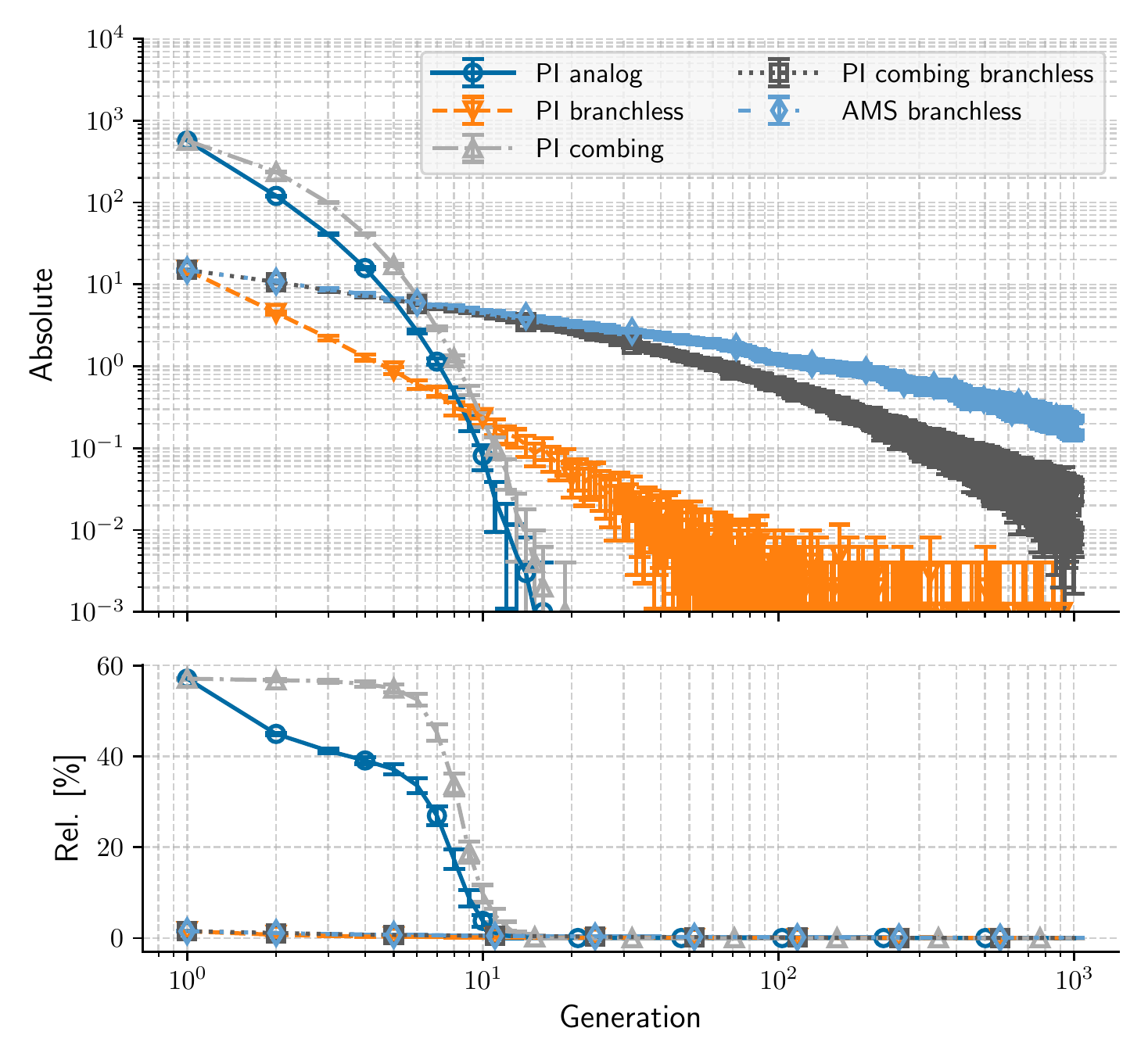}
  \caption{Absolute (top) and relative (bottom) mean number of independent neutron families killed by birth/death process over generations}
  \label{fig:families_killed_physics}
\end{figure*}

\begin{figure*}[!htb]
  \centering
  \includegraphics{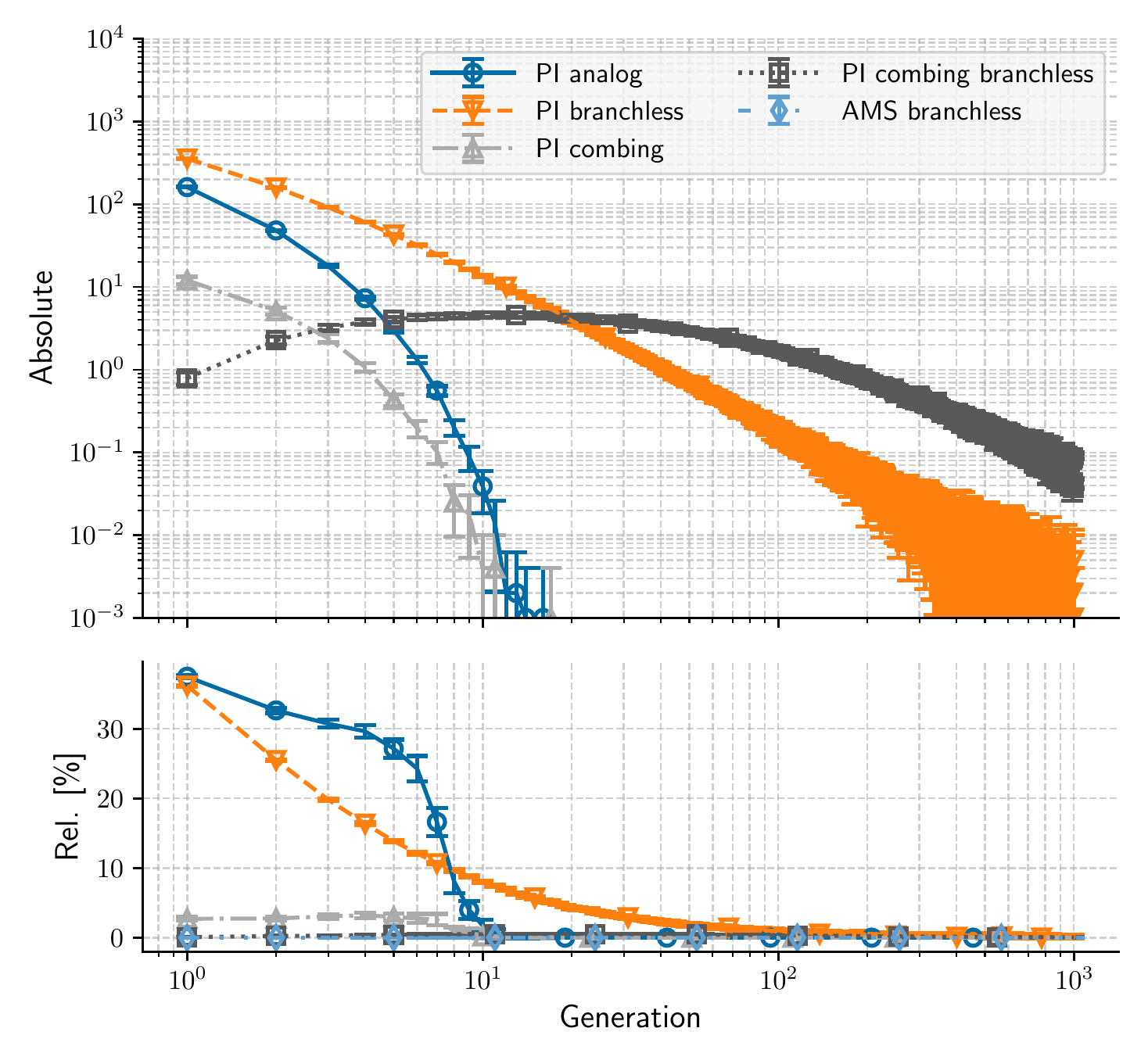}
  \caption{Absolute (top) and relative (bottom) mean number of independent neutron families killed by population control over generations}
  \label{fig:families_killed_pop_ctrl}
\end{figure*}

In the power iteration method, the death of a particle can occur from physical phenomena during the transport stage or by being combed out or not selected for duplication during the population control step.
To look at these two mechanisms in more detail, Figures \ref{fig:families_killed_physics} and \ref{fig:families_killed_pop_ctrl} present the number of independent families removed from the simulation throughout the transport stage and during the population control step, respectively.

Regarding the death occurring during transport, their relative number is higher when the branchless collision is not used, as seen in Figure \ref{fig:families_killed_physics}, because the method reduces the number of uncorrelated pairs that disappear during the transport step by preventing families from dying from capture.

Clusters of particles can form due to the asymmetry between particle death, which is likely to happen everywhere in the core, and the birth of new particles, which happen only at fission sites.
Regulating the particle population thus also affects the formation of clusters by removing independent families (which are not sampled during the population control step), reducing the number of uncorrelated particles in the process.
It also contributes to regrouping particles into areas of the geometry because the probability of sampling a particle from the fission bank in a region of the geometry will depend on the density of particles inside that region, hence favoring high-density regions like clusters.
In that regard, the combing is expected to be less "cluster-friendly" than sampling with replacement since particles can be sampled a limited number of times by combing.
As seen in Figure \ref{fig:families_killed_pop_ctrl}, the combing method shows excellent results with a meager killing rate, around a few percent. At the same time, the AMS does not appear in the absolute results (top figure) because the AMS does not remove particles during re-sampling.
It is because it never kills independent families since the "population control" operated during the re-sampling step, see Section \ref{sec:AMS_sampling_step}, only regenerates particles.
Obtaining the same effect without taking into account any importance function would eventually be possible by modeling the system as a Fleming-Viot particle system\footnote{To characterize the state of this system conditioned on its survival}.
Eventually, these observations are consistent with the findings of recent work on the role of population control on clustering in Monte Carlo iterated-fission-source calculations \cite{Sutton2022TowardClustering}.
It is interesting to notice that using branchless collision when population control is done by sampling with replacement (case PI branchless on Figure \ref{fig:families_killed_pop_ctrl}) causes more families to be terminated than the analog collisions (case PI analog).
This is because analog collisions induced more death during transport, so the total number of independent families is lower once the population control step is reached, as seen on the bottom plot of Figure \ref{fig:families_killed_pop_ctrl} presenting a relative number of families being killed in the case PI analog.

As a reminder, the number of particles per cycle was deliberately too low to enable the formation of neutron clusters to compare the methods regarding clustering issues.
In the end, both the AMS and the combing method helped reduce clustering effects by preserving more independent families if combined with the branchless collision.
In a production calculation, the number of particles would be higher to reduce the bias on the average value. However, for loosely coupled systems where this bias is limiting regarding the number of particles simulated in each generation, decreasing the required number of neutrons by using an appropriate method would be attractive regarding the global performances of the calculation.

\subsection{Variance estimation}\label{sec:variance_estimation}
\begin{figure*}[!htb]
  \centering
  \includegraphics{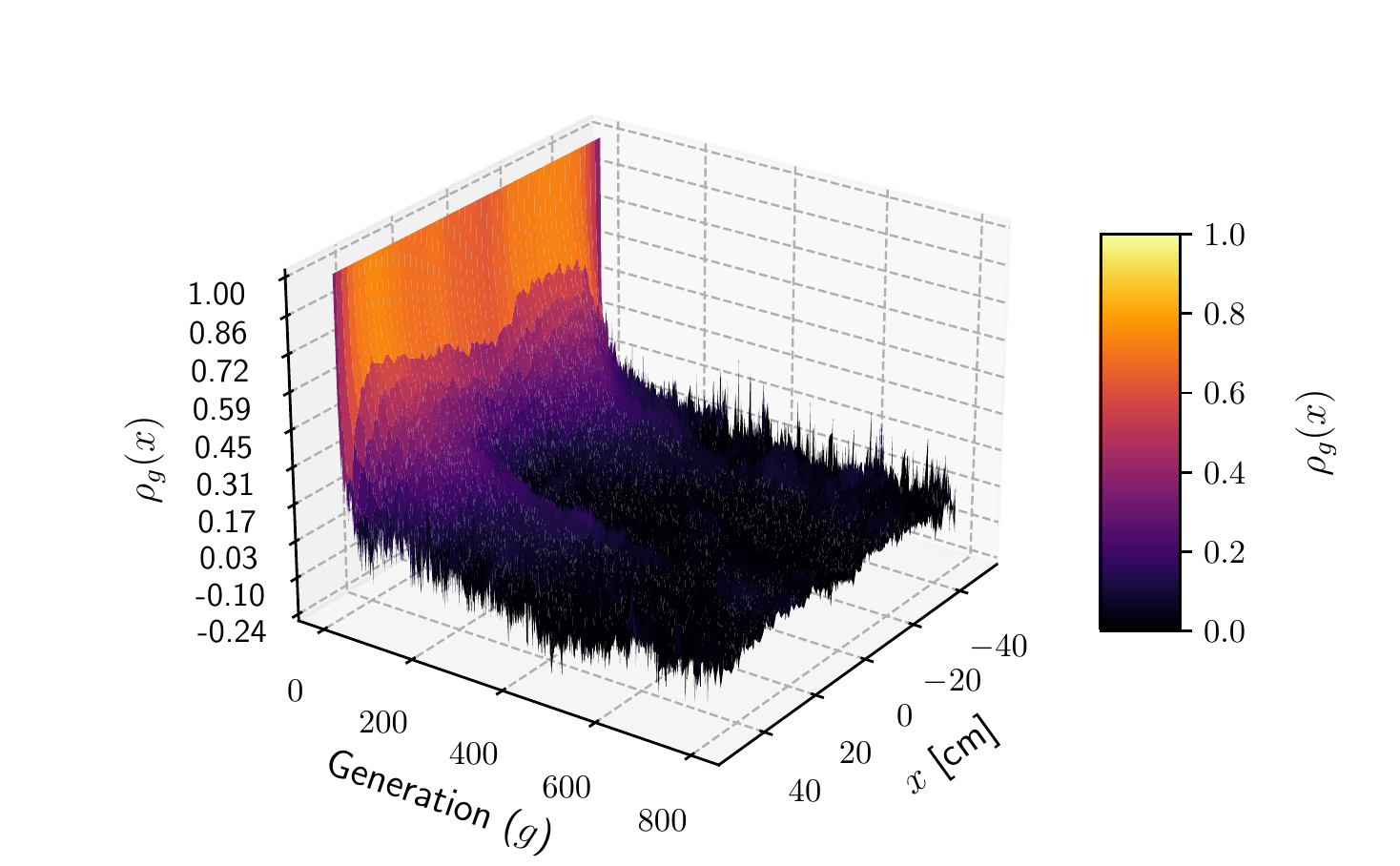}
  \caption{Cycle correlations for the PI analog case.}
  \label{fig:time_correl_PI_analog}
\end{figure*}

\begin{figure*}[!htb]
  \centering
  \includegraphics{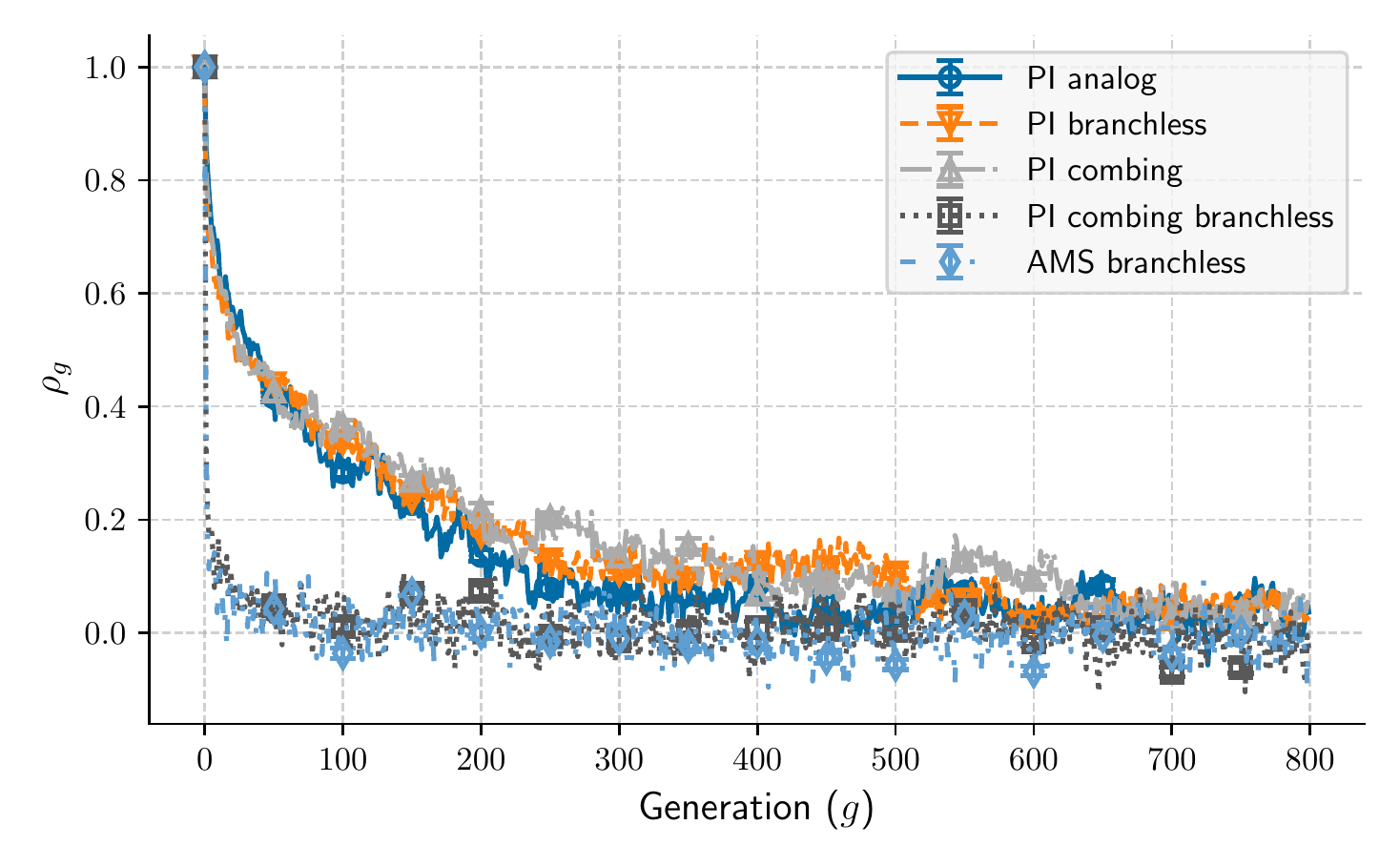}
  \caption{Cycle correlations for $x = 23.5$ cm.}
  \label{fig:time_correlations_1D}
\end{figure*}

Besides a bias on the mean flux estimates due to clustering, criticality calculations can also present a bias on variance estimates since scores are averaged over correlated generations.
In order to evaluate the bias on the flux variance estimation along the $x$-axis of the geometry, generational correlation coefficients were computed as a function of the neutron position along $x$.
Like spatial correlations, the generational correlations were estimated using Pearson's correlation coefficient 
\begin{equation}\label{eq:generational_correlations}
    \rho_g(x_l) = \frac{Cov\left[ \phi_0(x_l), \phi_g(x_l) \right]}{\sigma\left[ \phi_0(x_l) \right] \sigma\left[ \phi_g(x_l) \right]}
\end{equation}
where $\rho_g(x_l)$ is the correlation coefficient for the flux estimate in spatial bin $x_l$ between two generations $g$ apart, computed from $M$ independent simulations, $\phi_0(x_l)$ and $\phi_g(x_l)$ are flux estimates in spatial bin $x_l$ in the first and $k+1$-th active generations respectively, and $\sigma\left[ \phi_0(x_l) \right] \sigma\left[ \phi_g(x_l) \right]$ is the product of their standard deviation.
Figure \ref{fig:time_correl_PI_analog} illustrates the behavior of the generational correlations in different space bins in the case PI analog, which is expected to be the worst-case scenario regarding correlations.
In this figure, two local maxima appear along the $x$-axis (around -25 cm and 25 cm, which corresponds to $1/4$ and $3/4$ of the slab length), and three local minima around -50, 0 and 50 cm ($0$, $1/2$ and $1$ of the total length) in each generation.
This behavior has already been observed in previous work and is due to excitation of the eigenvector higher modes as explained in Ref. \cite{dumonteil2012automatic}.
To compare the different calculations, a slice along the $x$-axis is plotted in Figure \ref{fig:time_correlations_1D}, around $x = 25$ cm which is one of the locations where correlations are the strongest.
This figure highlights that generational correlations drop quickly to negligible levels when combing or AMS are combined with the branchless collision method.
In a nutshell, the real variance should be very close to the apparent one given by the Monte Carlo calculation in those two cases.
Computing the cycle correlations allows us to compute the real variance when estimating the Figure of Merit.

In order to compare the efficiency of the methods, the Figure of Merit (FoM) for the flux was computed in each spatial bin $x_l$ as such
\begin{equation}
    FoM(x_l) = \frac{1}{\sigma^2_{corr}(x_l) T_\textrm{calc}}
\end{equation}
where $T_\textrm{calc}$ is the calculation time and $\sigma^2_{corr}$ is the variance of the score which was computed accounting for generational correlations using Bienaymé's identity such that
\begin{equation}
    \sigma^2_{corr}(x_l) = \frac{\sigma^2(x_l)}{N}\left[1 + 2\sum_{g=1}^{G-1} \left( 1-\frac{g}{G} \right)\rho_k(x_l) \right]
\end{equation}
where $\sigma^2(x_l)$ is the variance between estimations of the flux in spatial bin $x_l$, $N$ is the size of the sample used to compute the average flux, $G$ is the number of active cycles and $\rho_k(x_l)$ is the generational correlations coefficients defined in Equation \ref{eq:generational_correlations}.

The computation times are presented in Table \ref{tab:calc_time}. 
All methods show similar orders of magnitude, with combing overall faster than the rest and AMS slightly slower. 
The combing speed is due to the way source neutrons are sampled, which is more efficient than sampling with replacement. 
Concerning the AMS, the transport part is slightly faster than in the power iteration due to the smaller number of collisions occurring in some generations (see Figure \ref{fig:entropy_AMS}). 
It appears that a nonnegligible amount of time is spent in the function that adds points into the AMS structure, which could probably be optimized.
\begin{table}[!htbp]
  \centering
\caption{Computation time [s] for each calculation}
    \label{tab:calc_time}
\begin{adjustbox}{width=1\textwidth} 
    \begin{tabular}{l c c c c c c }
        \toprule
        Case & Total & Transport & Sampling & Scoring & Sorting tracks & Adding points\\
        \midrule
        \multirow{2}{*}{PI analog} & \multirow{2}{*}{$1.822\times 10^4$} & $1.896\times 10^3$ & $7.954\times 10^3$ & $6.134\times 10^3$ & & \\
         &  & ($10$) \% & ($43$) \% & ($33$) \% & & \\
        \multirow{2}{*}{PI branchless} & \multirow{2}{*}{$1.779\times 10^4$} & $1.859\times 10^3$ & $7.805\times 10^3$ & $5.949\times 10^3$ & & \\
         &  & ($10$ \%) & ($43$ \%) & ($33$ \%) & & \\
        \multirow{2}{*}{PI combing} & \multirow{2}{*}{$1.497\times 10^4$} & $2.066\times 10^3$ & $4.377\times 10^3$ & $6.245\times 10^3$ & & \\
         & & ($13$ \%) & ($29$ \%) & ($41$ \%) & & \\
        \multirow{2}{*}{PI combing branchless} & \multirow{2}{*}{$1.392\times 10^4$} & $1.896\times 10^3$ & $3.990\times 10^3$ & $5.885\times 10^3$ & & \\
         &  & ($13$ \%) & ($28$ \%) & ($42$ \%) & & \\
        \multirow{2}{*}{AMS branchless} & \multirow{2}{*}{$2.094\times 10^4$} & $1.660\times 10^3$ & $4.166\times 10^1$ & $1.199\times 10^4$ & $3.638\times 10^1$ & $3.867\times 10^3$ \\
         & & ($7$ \%) & ($0$ \%) & ($57$ \%) & ($0$ \%) & ($18$ \%) \\
        \bottomrule
    \end{tabular}
  \end{adjustbox}
\end{table}

\begin{figure*}[!htb]
 \centering
  \includegraphics{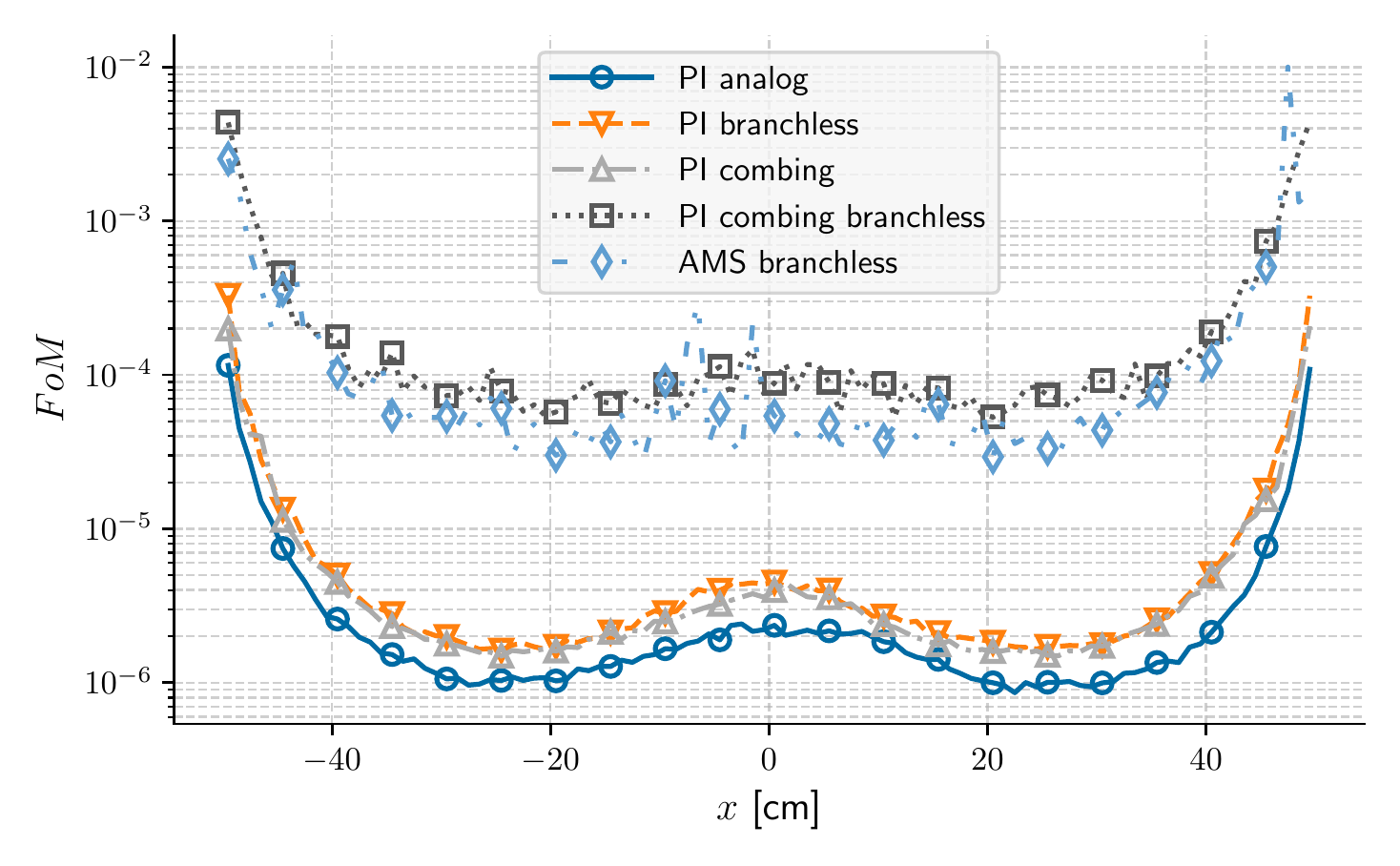}
 \caption{Figure of Merit for the flux estimation over $x$.}
 \label{fig:FoM}
\end{figure*}

The resulting FoM for the spatial flux is shown in Figure \ref{fig:FoM}.
Cases PI combing branchless and AMS branchless display better FoM, about one to two orders of magnitude more than the three other cases for all $x$.
Overall, preserving the maximum number of independent pairs of particles through appropriate population control, combined with the branchless collision method, which reduces the variance between fission chains length, improves the Figure of Merit of the spatial flux.
In that sense, the AMS seems to be an equivalent alternative to the combing method.

\FloatBarrier
\section{Conclusion}
In this work, the Adaptive Multilevel Splitting (AMS) algorithm initially designed for variance reduction and used in fixed source simulations has been extended to Monte Carlo neutron criticality calculations.
The results obtained with this methodology were compared to the ones obtained with the power iteration algorithm used in Monte Carlo calculations on a one-dimensional homogeneous reactor slab.
To assess for the population control impact on correlations and clustering, multiple population control methods were used in the power iteration.
The results produced by the different methods were compared regarding the average $k_\textrm{eff}$ and spatial flux, as well as spatial and generational correlations. 

Due to the fact that the AMS does not kill particles like usual population control techniques, it has allowed to highly reduce correlations levels.
It has been combined with the branchless collision method and has showed results almost identical to those obtained with the power iteration when the combing and the branchless collision methods were used.
Compared to the other cases (power iteration with sampling with replacement and/or analog collisions), the AMS branchless and the combing branchless displayed spatial and generational correlation levels close to nil, resulting in almost no clustering, despite a low number of neutrons per generation.
Overall, we managed to compute a fundamental flux distribution using the AMS with branchless collisions with a Figure of Merit (FoM) multiplied by 100 compared to an elementary power iteration, thus close in magnitude to the power iteration using the combing and branchless collisions methods.

The importance function chosen in the numerical applications remained quite simple due to the nature of the system.
Indeed, we do not expect much improvement in the Figure of Merit by changing the spatial shape of the importance in one-speed homogeneous problems.
Modeling more complex systems with a non-trivial adjoint solution should be necessary to further characterize this method's behavior, especially for loosely coupled systems.
In these systems, neutrons would have difficulties reaching certain regions, thus making the effects of the importance function even more significant and potentially potentially improving the FoM compared to the combing used in combination with branchless collisions. 

Another opening for this work could be to investigate, on the contrary, approaches that totally eliminate the importance map.
Indeed, the idea of using the AMS in criticality simulations was to characterize the asymptotic behavior of a system (e.g., the $k_\textrm{eff}$ and the fundamental flux) conditioned to its survival.
This approach does, in essence, not require an importance function to rank tracks and push neutron histories through time.
It could be possible to get rid of this function by treating the system as a Fleming-Viot process, thus benefiting from the population control to regenerate particles without killing independent families.

Finally, and more importantly, since the AMS has been capable of computing a steady-state spatial flux distribution, regardless of the reactivity of the system, it should be conceivable to take a step further and use it to model transients in kinetics calculations.
The target detector would therefore be defined in specific time bins, e.g., one could be interested in reducing the variance of the power distribution during the power peak.
In order to achieve variance reduction in specific time bins, the importance function would have to account for particles position in time.
Hence it would be helpful to be able to compute a time-dependent adjoint flux.
Going from a time-independent adjoint flux to a time-dependent one would also allow taking delayed neutron precursors importance into account.
Indeed, AMS branches can also carry the particle type as a parameter, making it possible to use multiple importance functions depending on the particles nature.

\pagebreak
\section*{Acknowledgments}

The authors would like to thanks Benjamin Dechenaux for his useful remarks on the present article, as well as Tony Lelièvre for helpful discussions on the Adaptive Multilevel Splitting method.

\pagebreak
\bibliographystyle{style/ans_js}                                                                           
\bibliography{bibliography,referencesFromMendeleyEric}

\end{document}